\documentclass[twocolumn,floatfix,showpacs,preprintnumbers,amsmath,amssymb,prb,superscriptaddress]{revtex4}

\usepackage{graphicx}
\usepackage{dcolumn}
\usepackage{bm}

\begin{document}

\title{Monodisperse self-assembly in a model with protein-like interactions
}

\author{Alex W.~Wilber}
\affiliation{Physical and Theoretical Chemistry Laboratory,
  Department of Chemistry, University of Oxford, South Parks Road,
  Oxford, OX1 3QZ, United Kingdom}
\author{Jonathan P.~K.~Doye}
\thanks{Author for correspondence}
\affiliation{Physical and Theoretical Chemistry Laboratory,
  Department of Chemistry, University of Oxford, South Parks Road,
  Oxford, OX1 3QZ, United Kingdom}
\author{Ard A.~Louis}
\affiliation{Rudolf Peierls Centre for Theoretical Physics,
 University of Oxford, 1 Keble Road, Oxford, OX1 3NP, United Kingdom}
\author{Anna C.~F.~Lewis}
\affiliation{Physical and Theoretical Chemistry Laboratory,
  Department of Chemistry, University of Oxford, South Parks Road,
  Oxford, OX1 3QZ, United Kingdom}

\date{\today}

\begin{abstract}
We study the self-assembly behaviour of patchy particles with `protein-like' interactions that can be considered as a minimal model for the assembly
of viral capsids and other shell-like protein complexes. We 
thoroughly explore the thermodynamics and dynamics of self assembly as a 
function of the parameters of the model and find robust assembly of all target
structures considered. 
Optimal assembly occurs in the region of parameter space where a free energy 
barrier regulates the rate of nucleation, thus preventing the premature 
exhaustion of the supply of monomers that can lead to the formation of 
incomplete shells. The interactions also need to be specific enough to prevent
the assembly of malformed shells, but whilst maintaining kinetic accessibility.
Free-energy landscapes computed for our model have a funnel-like topography 
guiding the system to form the target structure, and show that the torsional 
component of the interparticle interactions prevents the formation of 
disordered aggregates that would otherwise act as kinetic traps.
\end{abstract}


\maketitle

\section{Introduction}
\label{sec:Intro}

The self-assembly of simple building blocks into larger, ordered structures is a ubiquitous process in biology, and holds great promise for applications
in materials science and nanotechnology.\cite{Whitesides02b,Goodsell04} 
In particular, proteins are the building blocks for a vast array of biological 
structures, including capsules, fibres, tubes, sheets, channels and 
motors. 
Here, we restrict our interest to those examples where the self-assembly is 
both one-component and monodisperse, 
i.e.\ the structures formed are of a specific 
size and made up of multiple copies of the same protein.
The archetypal examples of this type are icosahedral virus capsids, which are 
designed to encapsulate the genomic material of the virus.
The capsid proteins are produced in large quantities inside host cells, 
and need to assemble correctly and spontaneously on a biological time scale in 
order for the virus to propagate. 
While many capsid assembly processes rely on interactions between the capsid 
proteins and scaffolding proteins or nucleic
acids,\cite{Bancroft67,Prevelige93,Dokland99,Parent05,Parent06}
capsids for a number of viruses have been shown to assemble 
\emph{in vitro} in the absence of these 
molecules.\cite{Salunke89, Rombaut90, Zlotnick99, Zlotnick00, Casini04}

There are also a significant number of non-viral proteins that form 
shell-like homomeric complexes with high 
symmetry.\cite{Goodsell00,Levy06d,Janin08b}
For example, ferritin is a complex that is made up of 24 sub-units with 
octahedral symmetry that stores iron inside its central cavity,\cite{Ford84} and
dihydrolipoyl acetyltransferase can form a 60-unit dodecahedral complex 
that is at the core of the pyruvate dehydrogenase multienzyme complex.\cite{Izard99}
However, the self-assembly behaviour of such complexes has been much less 
studied than for virus capsids.

Clearly, establishing a good understanding of the nature of the self-assembly 
process of protein complexes and virus capsids, as well as being of fundamental
interest, will be of value to many potential biomedical and bionanotechnological
applications, e.g.\ in the design of drugs which interfere with capsid 
assembly\cite{Zlotnick03b,Prevelige98, Stray07} or the use of capsids as 
vehicles for the delivery of drugs or other agents.\cite{Cho08,Lewis06}
Finally, an understanding of the features that enable such robust assembly 
in these protein systems may be very valuable in designing synthetic 
self-assembling systems.\cite{Zhang03,Glotzer07b}

There is now a considerable body of work studying the assembly of virus capsids,
both experimentally\cite{Sorger86, Zlotnick00, Willits03, Johnson04, Casini04, Parent05, Parent06, Parent07, Hanslip06, Zlotnick08b, Tuma08}
and theoretically.\cite{Endres02,Zlotnick05_TheoreticalAspects,Endres05,Zlotnick07,Rapaport99,Rapaport04,Rapaport08,Zandi06,vanDerSchoot07,Twarock06,Zhang06,Schwartz08,Hagan06,Hagan08,Hagan08_2,Brooks07,Brooks08,Nguyen08b,Wales05,Fejer09,Glotzer07}
The transient nature of the intermediates present a particular challenge to the
experiments. By contrast, characterizing species with short lift-times is much 
less of an obstacle for simulation and modelling,
and so they can play an important and complementary role by
illuminating the mechanisms of assembly. Instead, the problem for simulations
is the wide range of time and length scales associated with capsid assembly. 
For example, although all-atom simulations of small viruses are now 
possible,\cite{Freddolino06} the time scales associated with self-assembly are 
far beyond what is currently feasible. Therefore, it is necessary to use much 
simpler coarse-grained models, the hope being
that if self assembly has general rules as to what conditions and subunit 
designs lead to successful assembly, these should be accessible through such 
models.

In this vein, a number of computational approaches have been directed
towards capsid assembly. Kinetic models consider the populations of particular 
clusters and model cluster growth by assigning rate constants to 
combination/breakup events, using either a differential 
equation\cite{Endres02,Endres05} or a discrete event 
approach.\cite{Zhang06,Schwartz08}
However, these models do not include any information about the spatial 
positions of subunits, and only consider a limited set of assembly pathways.
Simulations do not suffer from these restrictions, but it is more challenging
to generate good statistics and thoroughly analyse the parameter space.
Indeed, it is only relatively recently that models with fully reversible 
dynamics have been studied.
The simulations can be grouped into two types depending on whether the protein
subunits are modelled as anisotropic objects consisting of a rigid structure 
of smaller spheres\cite{Rapaport04,Rapaport08,Brooks07,Brooks08,Nguyen08b} 
or as spherical particles with anisotropic `patchy' interactions.\cite{Hagan06}
The latter approach is typically computationally less 
intense, and is the one we use here.

Previously,\cite{Wilber07} we have used such a patchy particle model to study 
the assembly of icosahedral clusters, and in the accompanying 
paper this study is extended to the remainder of the Platonic 
solids.\cite{accompanying}
However, in these studies the model included no constraints 
on the torsional (dihedral) angle between interacting particles. While
this is likely to be appropriate for synthetic `patchy' colloids and nanoparticles,\cite{Glotzer07b,Cho07} it does not represent protein-protein interactions well since the complex structure
of the interfaces between the proteins will tend to restrict rotation. 
Such torsional constraints are also relevant to recently created 3- and 
5-armed DNA building blocks that can assemble into tetrahedra, 
dodecahedra, truncated icosahedra,\cite{He08} cubes\cite{Zhang09} and 
icosahedra,\cite{Zhang08} and where the relative orientation of these units
is controlled by the number of turns of the DNA double helix along each arm.

Here we consider the effects of adding a torsional component to the
interparticle potential of this minimal model, and proceed to map out the 
behaviour of our systems over a wide region 
of parameter space. We examine the dynamics and mechanisms of assembly, and 
consider the behaviour of the system in regions in which assembly is not 
successful in order to understand those processes which compete with 
successful assembly. 
This work has a number of distinctives compared to previous simulation studies 
of viral assembly. 
Firstly, we do not just consider the assembly of an icosahedral target that is 
equivalent to a $T=1$ capsid, but more
generally consider structures with other symmetries that are also relevant to
non-viral protein complexes.
Secondly, the model is simpler than those previously studied, thus allowing
us to survey the behaviour of the model very comprehensively. In particular,
we are able to connect the dynamics to the underlying thermodynamics of the 
model, and obtain detailed pictures of the free energy landscapes governing 
assembly.
Finally, comparison to the non-torsional model considered previously,
allows us to understand what features of the behaviour arise because of this
component in the interparticle interactions, 
and such information will be particularly relevant for those trying to 
design synthetic particles that are able to self-assemble.

\section{Methods}

\subsection{Model}

The model system we use is a slightly modified version of that used previously\cite{Wilber07,Doye07,Noya07b} and in the accompanying paper.\cite{accompanying}
It has also recently been used to study the assembly of tetrameric protein complexes.\cite{Villar09}
The model consists of spherical particles with a number of sticky patches, which are defined by patch vectors. The interaction potential is pairwise and is
based on the Lennard-Jones potential,
\begin{equation}\label{eqn:LJ} V_{\rm LJ}(r) = 4\epsilon\left[ \left( \frac{\sigma_{\rm LJ}}{r}
    \right)^{12} - \left( \frac{\sigma_{\rm LJ}}{r} \right)^{6} \right], \end{equation}
but the attraction is modulated by a pair of orientationally dependent terms,
$V_{\rm ang}$ and $V_{\rm tor}$. $V_{\rm ang}$ is a function of how directly the two patches are pointing at one another, while the additional
factor $V_{\rm tor}$ is a function of the torsional angle between the two particles, i.e.\ it varies as one of the particles is rotated about the
vector connecting the two particles. Thus, the complete potential is
\begin{eqnarray}
\label{eqn:potential}
\lefteqn{V_{ij}({\mathbf r_{ij}},{\mathbf \Omega_i},{\mathbf \Omega_j}) = }
\\
 & & \left\{
    \begin{array}{ll}
       V_{\rm LJ}(r_{ij}) & r<\sigma_{\rm LJ} \nonumber \\ \nonumber
       V_{\rm LJ}(r_{ij})
       V_{\rm ang}({\mathbf {\hat r}_{ij}},{\mathbf \Omega_i},{\mathbf \Omega_j})
       V_{\rm tor}({\mathbf {\hat r}_{ij}},{\mathbf \Omega_i},{\mathbf \Omega_j})
                       & r\ge \sigma_{\rm LJ}, \end{array} \right.
\end{eqnarray}
where ${\mathbf \Omega_i}$ is the orientation of particle $i$. For any pair of
particles only the pair of patches that maximizes $V_{\rm ang}V_{\rm tor}$ are
considered to interact.
$V_{\rm ang}$ has the form:
\begin{eqnarray}
V_{\rm ang}({\mathbf {\hat r}_{ij}},{\mathbf \Omega_i},{\mathbf \Omega_j})&=&
G_{ij}({\mathbf {\hat r}_{ij}},{\mathbf \Omega_i})
G_{ji}({\mathbf {\hat r}_{ji}},{\mathbf \Omega_j})\\
\label{eqn:AngMod}
G_{ij}({\mathbf {\hat r}_{ij}},{\mathbf \Omega_i})&=&
\exp\left(-\frac{\theta_{kij}^2}{2\sigma_{\rm ang}^2}\right),
\end{eqnarray}
where $\sigma_{\rm ang}$
is a measure of the width of the Gaussian,
$\theta_{kij}$ is the angle between patch vector $k$ on particle $i$
and the interparticle vector $\mathbf r_{ij}$.

In the calculation of $V_{\rm tor}$, an additional reference vector on each particle needs to be defined. We have chosen this reference
vector in each case to lie on the symmetry axis of the particle. $V_{\rm tor}$ is a
maximum when the projections of the two reference vectors onto the plane perpendicular to the interparticle vector lie parallel. Specifically, if
$\phi$ is the angle between the projections, then
\begin{equation}
V_{\rm tor} = \exp\left(-\frac{\phi^2}{2\sigma_{\rm tor}^2}\right)
\end{equation}
where $\sigma_{\rm tor}$ is a measure of the width of the Gaussian. 
The effect of the inclusion of $V_{\rm tor}$ in the potential is to penalise twisting around the interparticle vector, with smaller
values of $\sigma_{\rm tor}$ giving a stronger constraint.
In order to reduce the number of parameters to consider 
and because we consider it physically reasonable that $\sigma_{\rm tor}$ and $\sigma_{\rm ang}$ are coupled, 
we set $\sigma_{\rm tor} = 2\,\sigma_{\rm ang}$ throughout this paper,
so that the specificity of interactions is given only in 
terms of $\sigma_{\rm ang}$. 
Our results are relatively insensitive to the precise value of the ratio of these parameters.

The patch vectors are chosen such that they point directly at the neighbouring
particles in the target structure. 
This choice, along with that of the reference vector above, ensures that the 
target structures are lowest in energy for sufficiently specific patches.
The target clusters that we consider are the five Platonic solids. Thus,
each particle has internal $C_n$ symmetry because of the presence of $n$ 
equivalent and regularly-arranged patches. 
Therefore, when comparing to virus capsids and other large homomeric 
protein complexes, the particles do not represent 
individual proteins, which have no symmetry, but cyclic protein complexes
that can further assemble into larger complexes. 
For example, the icosahedron-forming particles could be considered to 
represent the pentameric capsomers often posited as intermediates in viral 
capsid assembly\cite{Willits03,Hanslip06} and the complete icosahedron a $T=1$ capsid. 
Similarly, the dodecahedron-forming particles could represent
dihydrolipoyl acetyltransferase trimers.\cite{Izard99}
In this context, 
we are modelling the second stage of a hierarchical self-assembly 
process.\cite{Levy08}

For computational efficiency we use a cut-and-shifted version of the potential 
with a cutoff distance of $3\,\sigma_{\rm LJ}$, and also shift the crossover 
distance in Eq.\ \ref{eqn:potential} so that it occurs when the cut-and-shifted
potential passes through zero.

\subsection{Simulation}
\label{sec:Simulation}

All our simulations are based on Metropolis Monte Carlo simulations in the canonical ensemble. We
restrict the translational and rotational moves to be local, so that the particle motion mimics the diffusive behaviour of proteins and nanoparticles in
solution. Where equilibrium statistics are needed we make use of the umbrella sampling technique, applying an order parameter dependent bias to facilitate
the crossing of free energy barriers. For more details see the accompanying paper.\cite{Wilber07}

A particle number density of $0.15\,\sigma_{\rm LJ}^{-3}$ is used for all simulations. This represents a significantly higher concentration than would
normally be considered in \emph{in vitro} studies involving proteins. Because 
the time scales accessible to simulation are far shorter than those available 
to experiment, it is necessary to use higher concentrations in order to 
observe assembly. 
For our original model without torsional constraints we found that for 
concentrations in the range $0.025-0.4\,\sigma_{\rm LJ}^{-3}$ 
the final yields only show a weak dependence of on concentration.\cite{Wilber07}

\begin{figure}
\includegraphics[width=6.5cm]{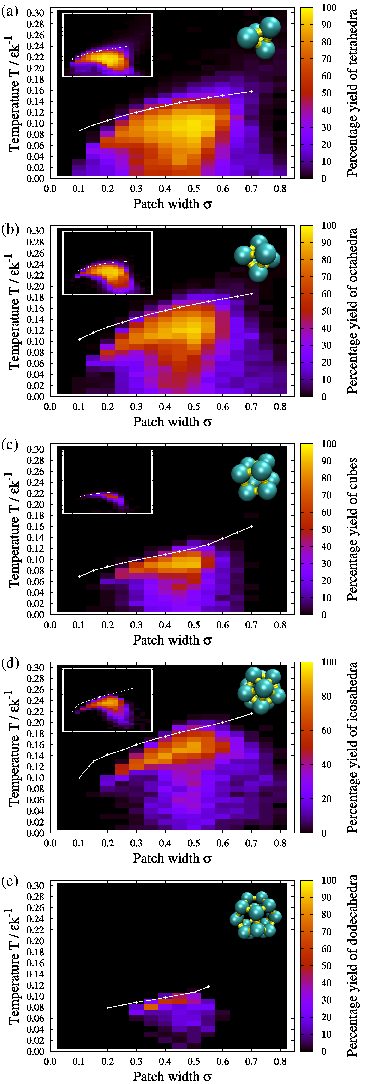}
\caption{\label{fig:AllYields}(Colour Online) The percentage yield of target clusters formed after $80\,000$ MC cycles
as a function of the patch width $\sigma_{\rm ang}$ (measured in radians) and temperature at a number density of $0.15\,\sigma_{\rm LJ}^{-3}$ for systems of
1200 particles designed to form (a) tetrahedra, (b) octahedra, (c) cubes, (d) icosahedra and (e) dodecahedra. 
The insets show the equivalent plots for simulations without torsional 
constraints, where the axes span the same parameter ranges as the
main plots. No inset is included for dodecahedra since without torsional constraints there was no conditions under which dodecahedra assembled.
The images in the top right of each plot show the relevant target structure.}
\end{figure}

\section{Results}

\begin{figure}
\includegraphics[width=7.1cm]{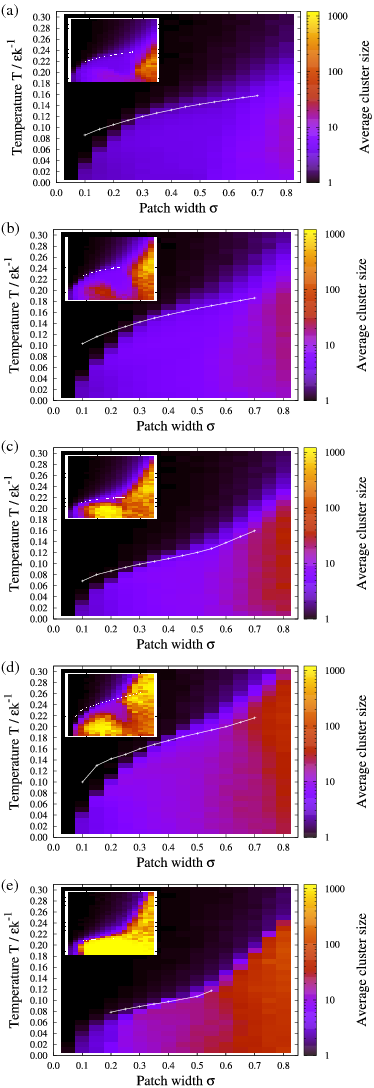}
\caption{\label{fig:AllSizes}(Colour Online) The mean cluster size 
(averaged over particles) of systems designed to form (a) tetrahedra, 
(b) octahedra, (c) cubes, (d) icosahedra and (e) dodecahedra, for the same
simulations as Fig.\ \ref{fig:AllYields}. 
The insets show the equivalent plots for simulations without torsional constraints, where the axes span the same parameter ranges as the main plots.}
\end{figure}

In order to map the behaviour of our systems over a wide region of parameter space, we performed large arrays of simulations
with varying values of the patch width $\sigma_\mathrm{ang}$ and the temperature $T$, for each of the Platonic solids. The results are shown in
Figs.\ \ref{fig:AllYields} and \ref{fig:AllSizes}. Note that, although we will
typically talk about the dependence of the behaviour on temperature, we could 
equivalent have used the interaction strength, 
where decreasing temperature corresponds to increasing interaction strength.
In Fig.\ \ref{fig:AllYields}, which gives the yields of successfully assembled clusters, a region of high yield is visible for each target structure at
moderate values of temperature and patch width, where the target is 
thermodynamically stable and kinetically accessible. 
Moving away from this region, yields fall away as competing processes 
become dominant.

The high temperature region corresponds to a gas of monomers and small clusters. Above a certain temperature, which we term $T_c$, the target clusters
cease to be stable with respect to this high entropy gas. This temperature is a strong function of the patch width $\sigma_\mathrm{ang}$ since wider patches
lead to higher entropies for the target clusters, arising from internal vibrations. Fig.\ \ref{fig:AllSizes}, which shows the average cluster
sizes achieved as a function of patch width and temperature, clearly shows the extent of the monomer gas region.

In the early stages of the simulations below $T_c$, 
the particles tend to rapidly assemble into curved shell-like structures. 
The inclusion of torsional constraints
in the potential favours a certain curvature in the structures formed, such that even erroneous structures are typically still hollow shells of roughly the
correct size
and shape. In particular the constraints tend to enforce convexity on the structures, greatly restricting the range of possible structures which can
compete with the target structure.

\begin{figure*}
\includegraphics[width=14.7cm]{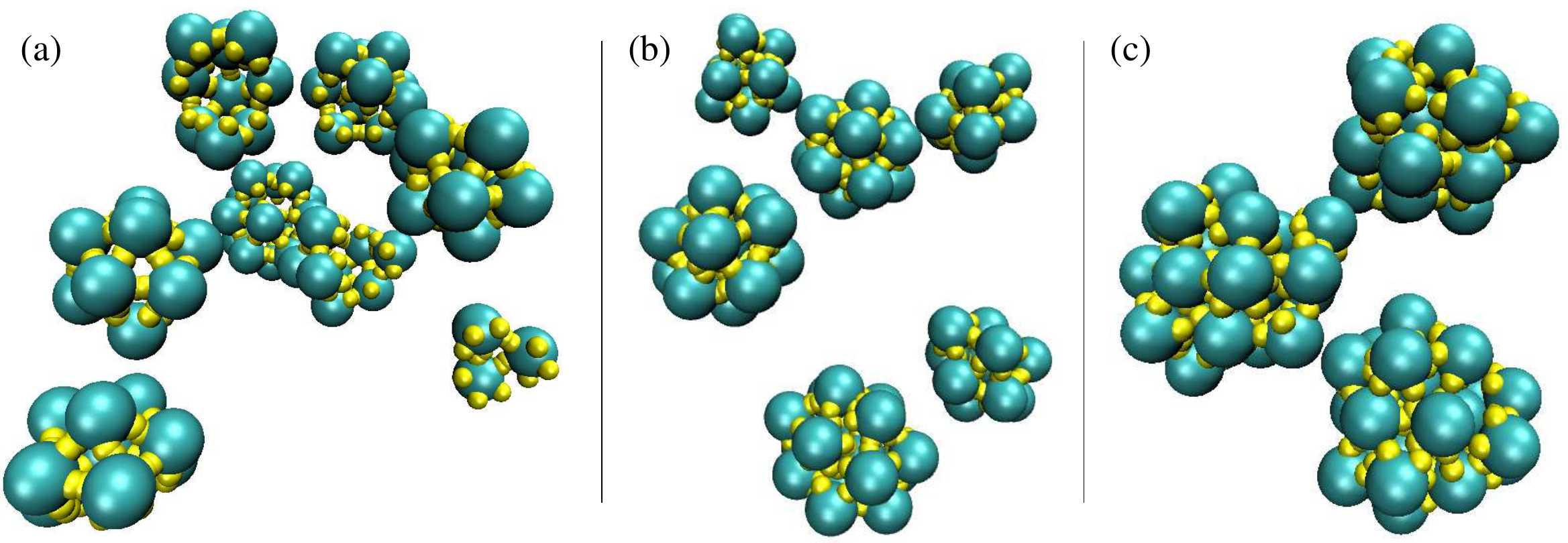}
\caption{\label{fig:IcosaTorSnaps}(Colour Online) Snapshots of typical configurations of 72 icosahedron-forming particles.
(a) $T=0.08, \sigma_{\rm ang} = 0.35$, showing monomer starvation, (b) $T=0.08, \sigma_{\rm ang} = 0.6$,
showing misformed clusters, and (c) $T=0.16, \sigma_{\rm ang} = 0.75$, showing liquid droplets.}
\end{figure*}

At high values of the patch width $\sigma_\mathrm{ang}$, the interactions between particles become less specific. Low energies can still be obtained
by significantly distorted clusters, and so with larger $\sigma_\mathrm{ang}$ increasingly malformed clusters are observed, until correctly formed
clusters become a rarity. These malformed clusters are somewhat similar to the 
monster particles which have been observed experimentally in 
Turnip Crinkle Virus,\cite{Sorger86} and in simulations
by Nguyen \emph{et al.}.\cite{Brooks07} However, the nature of our interaction potential limits the range of structures which can be formed, so that the
fused-shell structures seen in these papers are not seen. Instead we find convex structures similar to those found in more recent work by Nguyen and
Brooks.\cite{Brooks08} At still higher values of $\sigma_\mathrm{ang}$ the patches are sufficiently wide that they lose their specificity, and
instead of forming hollow shells, the system forms large, roughly spherical 
liquid-like droplets.

At low temperatures more complicated behaviour is observed, with generally lower yields but some successful assembly even at very low temperatures for
moderate values of $\sigma_{\rm ang}$. The low yields arise because rapid nucleation leads to the removal of almost all the monomers from the system by the
growing clusters, leaving many clusters half-formed and unable to grow further. This phenomenon of monomer starvation is a recurring theme of virus
assembly, and has been reported in a number of previous experimental 
studies,\cite{Parent05,Parent06} as well as being predicted by
kinetic models\cite{Endres02,Zhang06} and simulations.\cite{Hagan06,Rapaport04,Rapaport08,Brooks07}
We shall return later to consider the reason for the recovery in yield at moderate patch widths.

Fig.\ \ref{fig:IcosaTorSnaps} shows typical configurations for small systems in different regions of parameter space for the icosahedron-forming particles. 
Fig.\ \ref{fig:IcosaTorSnaps}(a)
shows a system displaying monomer starvation. Six partially formed clusters are visible, as well as one trimer and one 14-particle cluster which
is not easily able to rearrange because of the low temperature and patch width. This system is unlikely to form any correctly formed clusters within
accessible timescales. Fig.\ \ref{fig:IcosaTorSnaps}(b) shows a system at higher patch widths, such that misformed clusters are relatively low in energy
and are frequently observed. Two correctly formed icosahedra are visible, along with a number of misformed clusters. Finally,
Fig.\ \ref{fig:IcosaTorSnaps}(c)
shows liquid droplets formed at very high $\sigma_{\rm ang}$. Unlike the misformed clusters seen in Fig.\ \ref{fig:IcosaTorSnaps}(b), these droplets are
not hollow, and the particles are mobile within the droplets.

The general form of the behaviour of the systems is to first order independent 
of the target structure. The plots for each of the targets in
Figs.\ \ref{fig:AllYields} and \ref{fig:AllSizes} have essentially the same form in each case. However, some differences are also observed. Most strikingly,
the temperature delineating the transition from target clusters to a monomer gas, $T_c$, varies considerably. In general $T_c$ is greater
for smaller clusters (since there is a lower entropic cost for the formation of these clusters), and also greater for higher numbers of patches per
particle (since the clusters are more energetically favoured). These effects lead to fairly similar values of $T_c$ for the tetrahedron, octahedron and
icosahedron, where the differences from the two effects largely cancel, and lower values for the cube and the dodecahedron.

The insets in Figs.\ \ref{fig:AllYields} and \ref{fig:AllSizes} show the results for equivalent systems without torsional constraints. The most striking
differences are visible in Fig.\ \ref{fig:AllSizes}, where the insets show that in the absence of torsional constraints, extremely large clusters are
formed over wide ranges of parameter space. These large clusters correspond to disordered kinetic (at low $T$) or thermodynamic (at high 
$\sigma_\mathrm{ang}$) aggregates. 
Torsional constraints prevent the formation of these large and disordered
aggregates by enforcing convexity, the only exception being at the largest 
values of $\sigma_{\rm ang}$ in Fig.\ \ref{fig:AllSizes} that are well away from
the region of successful assembly.

For systems without torsional constraints, 
the competition between aggregation and self-assembly can severely limit 
the yield of the target structure.\cite{accompanying}
The effects of this competition 
are particularly noticeable for cubes, where assembly is limited to a 
small region, and dodecahedra, where assembly does not occur at all. 
For these two targets the yields dramatically increase
when torsional constraints are included because the competing aggregates
cannot form.

For the targets that readily assemble with and without torsional constraints, there is one region in which the yield is clearly decreased by the inclusion of torsional constraints. At patch widths slightly higher and temperatures slightly
lower than the optimum, the systems without torsional constraints assemble by a ``budding mechanism''.\cite{Wilber07}  
In this mechanism the particles first coalesce into
disordered aggregates, which then rearrange and bud off completed clusters. Since torsional constraints prevent the formation of large aggregates, they have the
effect of inhibiting this indirect assembly mechanism.

These effects are clearly visible in the different shapes of the regions of successful assembly with and without
torsional constraints in Fig.\ \ref{fig:AllYields}. The ``lobe'' of successful assembly due to budding is missing in the simulations with torsional
constraints, while a new region of partially successful assembly arises at low temperature and moderate patch widths, in part due to the lack of competition 
with large kinetic aggregates.

\begin{figure}
\includegraphics[width=8.4cm]{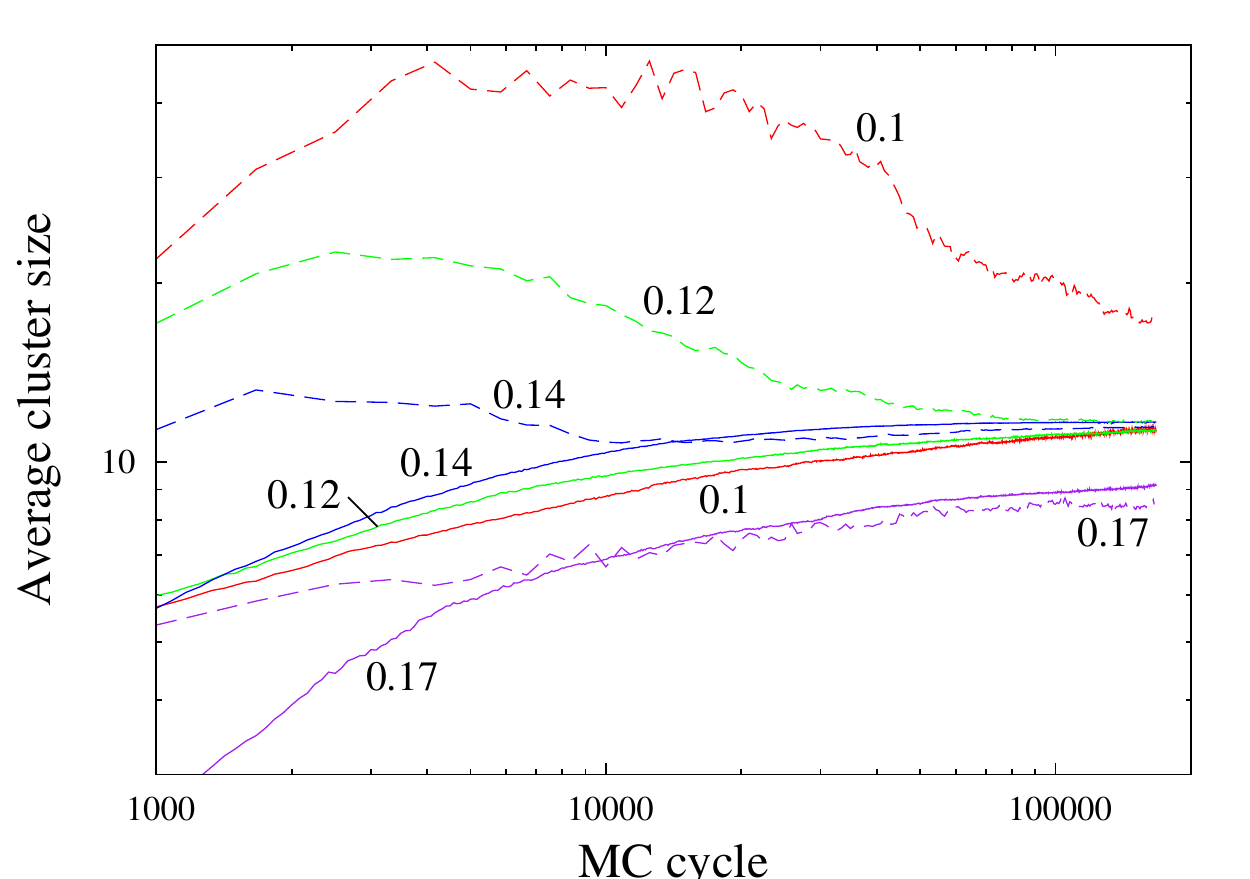}
\caption{\label{fig:AvSizes}(Colour Online) The mean cluster size (sampled over particles) as a function of number of MC cycles at $\sigma_{\rm ang}=0.45$
and at different temperatures (as labeled) for a system of 1200 icosahedron-forming particles. 
The dashed lines show the equivalent results for systems 
without torsional constraints. 
Each line is an average over ten simulations.}
\end{figure}

Assembly in systems with torsional constraints proceeds exclusively by the stepwise addition of monomers and small clusters
onto growing clusters. 
Fig.\ \ref{fig:AvSizes} shows the average cluster size as a function of time at different temperatures, with and without torsional
constraints. While in the absence of torsional constraints the average cluster size sometimes approaches the target size from above (indicating that
assembly proceeds by the budding mechanism), the average sizes for the simulations which include torsional constraints all approach the target size from
below, consistent with a stepwise growth mechanism.

\begin{figure*}
\includegraphics[width=16cm]{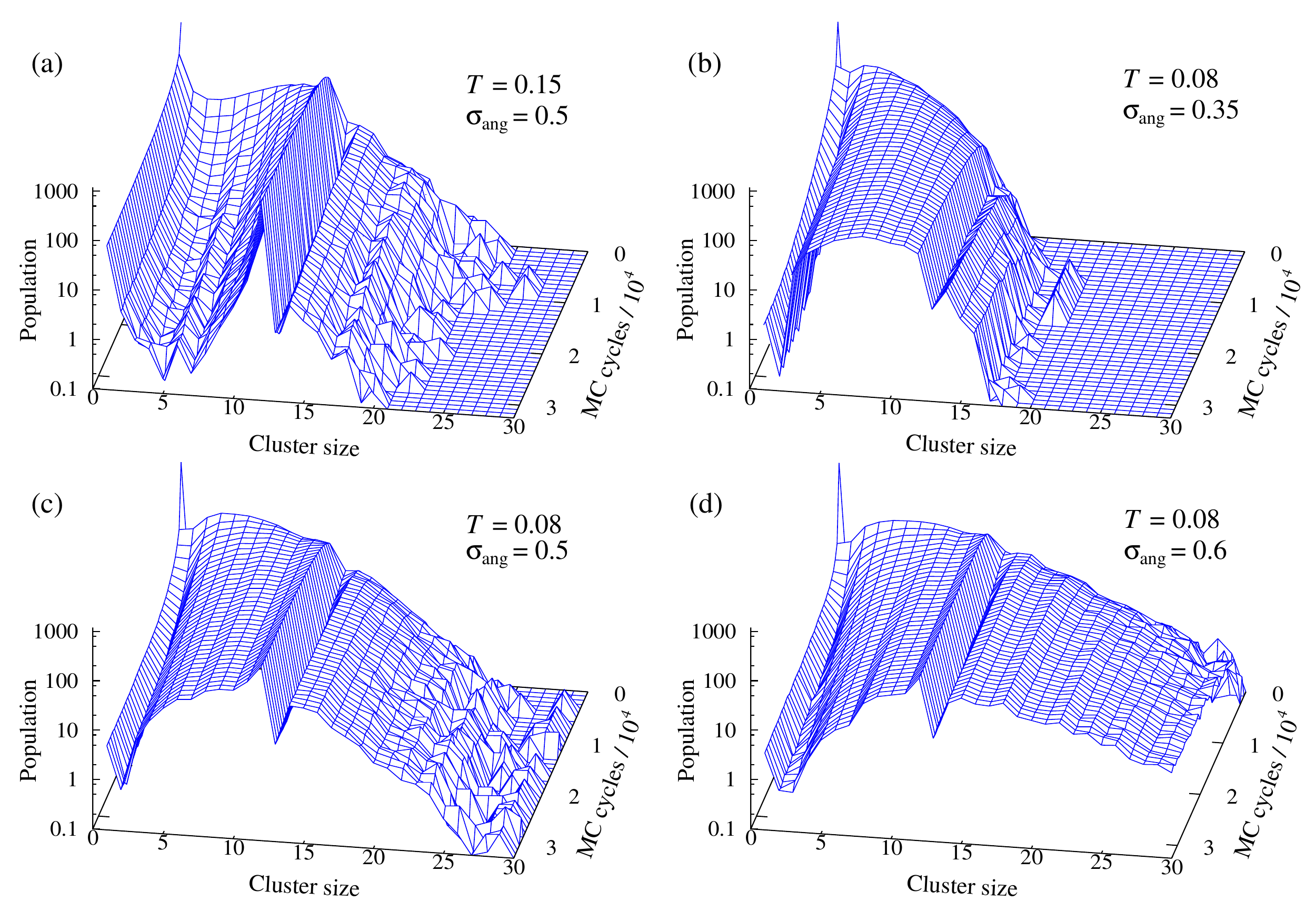}
\caption{\label{fig:ArdPlots}(Colour Online) The populations of particles in 
clusters of different sizes as a function of number of MC cycles for systems 
of 1200 icosahedron-forming particles, averaged over 100 repetitions.
The temperatures and patch widths are as labelled,
such that (a) is at the optimal conditions for assembly, (b) is in the region of monomer starvation, (c) shows recovery at moderate patch widths, and (d)
shows a loss of specificity at high patch widths. A log scale is used so that the very small populations of intermediates are discernible. States with
zero population have been set to a minimum value of 0.1 to aid in visualisation.
}
\end{figure*}

Figure \ref{fig:ArdPlots} provides a detailed picture of the dynamics by 
showing the numbers of particles in clusters of a given size as a function of 
time, under four different sets of conditions for the example 
of the icosahedron. Under optimal assembly conditions 
(Fig.\ \ref{fig:ArdPlots}(a)), there is only a small
population of intermediate-sized clusters and as time progresses the population
of monomers falls off slowly as more complete icosahedra are formed. That
the population of intermediate-sized clusters is always less than monomers
or icosahedra (after an initial lag time required for icosahedra to start
forming) is indicative of a significant free energy barrier for the formation 
of the target structure. As has been noted 
previously,\cite{Parent05,Parent06,Endres02,Zhang06} 
relatively slow nucleation is a prerequisite for successful capsid assembly.
When few nuclei are formed, the population of monomers remains high for longer 
allowing the majority of the nuclei to grow to completion. 

At lower temperatures, such as in Fig.\ \ref{fig:ArdPlots}(b), the effect of 
monomer starvation is evident. Under these conditions, the formation and growth
of cluster is all downhill in free energy. The consequent rapid nucleation of 
clusters leads to a rapid decrease in the population of monomers, and many
of the partially formed clusters are unable to grow to completion.
At this point assembly is effectively stalled, and can only proceed through a slow process of breakup of the existing clusters. This is the reason for the low yields observed at low values of $T$ and $\sigma_{\rm ang}$.

For these combination of reasons, optimal assembly occurs fairly close to $T_c$.
This result is consistent with Zlotnick's assertion that ``weak'' bonds are 
sufficient and indeed optimal for efficient assembly of 
viruses\cite{Zlotnick03,Zlotnick07} 
(but of course while still being strong enough that the capsids are 
stable).\footnote{It is noteworthy that in the work of Schwartz and coworkers
\cite{Zhang06,Schwartz08} the final yield is found to increase monotonically 
with decreasing binding rate constant (analogous to increasing temperature).
This unphysical behaviour results from some irreversible steps in their kinetic
scheme that means once formed the target structure can never disassemble.}

Although the phenomenon of monomer starvation is found at low temperature regardless of $\sigma_\mathrm{ang}$, yields are found to recover for moderate
patch widths. This recovery is observed in Fig.\ \ref{fig:ArdPlots}(c) 
and is due to an interesting mechanism of combination and rearrangement of clusters. Pairs of partly formed clusters
occasionally collide and stick together. Most often the two parts will not fit perfectly to form a complete cluster, and so a process of rearrangement
then occurs. If excess particles are present, they may often be ejected from the rearranging cluster.
Even closed shells are able to take advantage of this mechanism if they contain too few particles. If a small cluster approaches a strained region of
a shell it may be able to insert itself to form a complete target cluster; again, any unneeded particles may be ejected. These events are readily
visible in movies of our simulations. As an example of a typical event, a ten-particle closed cluster was observed to encounter a three-particle triangle.
The triangle
became loosely bound to the outside of the closed shell, then over time the particles in the shell moved apart such that two particles from the triangle
were able to incorporate into the shell to form a perfect icosahedron. The bonds with the remaining particle were broken, and it diffused away.

These processes depend on $\sigma_\mathrm{ang}$ taking at least a moderate value for several reasons. Firstly, collisions between partly formed clusters
are much more likely to result in binding interactions when the patches are wider. Secondly, insertion of small clusters into closed shells only becomes
possible when the patches are sufficiently wide that both the inserting 
fragment can become attached to the outside of the shell, 
despite the highly suboptimal angles between the respective patches, and 
the distortion required for insertion is feasible. 
Most importantly, the energy barriers for rearrangement are greatly decreased with wide patches, since the intermediate states can
be stabilised by interactions between poorly aligned patches. These mechanisms of combination and rearrangement lead, for moderate values of
$\sigma_\mathrm{ang}$, to reasonable yields at long times even at very low temperatures.

For even higher values of $\sigma_\mathrm{ang}$, yields fall off once more, and 
the kinetics in this regime are illustrated in Fig.\ \ref{fig:ArdPlots}(d). 
While the mechanisms of combination and rearrangement remain effective 
(hence the number of small intermediates again falls off with time),
the system is no longer so strongly driven to form correctly assembled clusters, since mildly misformed clusters will have similar energies per particle to correctly
formed clusters. Excess particles will only rarely be ejected, and many of the clusters formed are oversized.

\begin{figure}
\includegraphics[width=6.0cm]{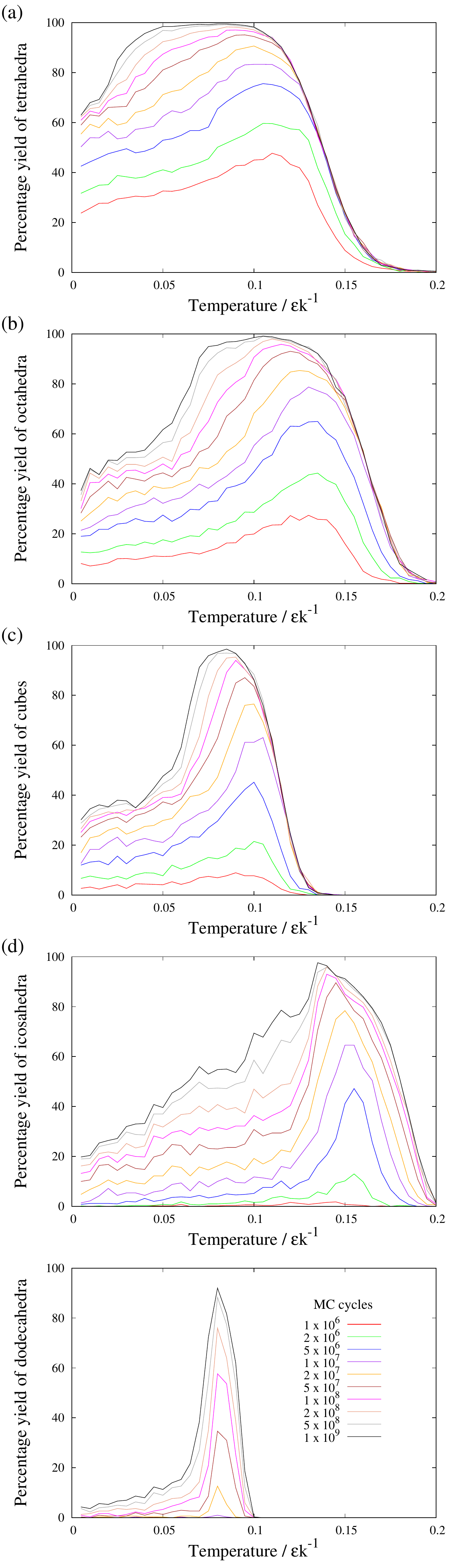}
\caption{\label{fig:YieldsWTime}(Colour Online) The yields of
(a) tetrahedra, (b) octahedra, (c) cubes, (d) icosahedra and (e) dodecahedra after different numbers of simulation steps as a function of temperature $T$,
in simulations of systems of 1200 particles. In each case the value of $\sigma_\mathrm{ang}$ was chosen
where maximum yields were obtained in Fig.\ \ref{fig:AllYields}, giving $\sigma_\mathrm{ang} = 0.5$ except in the case of the dodecahedra, where
$\sigma_\mathrm{ang} = 0.35$. Each data point is an average over five simulations. All the plots use the same key, inset in the plot
for dodecahedra.}
\end{figure}

Fig.\ \ref{fig:YieldsWTime} provides an alternative perspective on many of the effects discussed above. It shows the yields as a function of temperature
and time for each of the
shapes, at the optimal value of $\sigma_{\rm ang}$ in each case. A number of effects are visible. Considering first the plot for icosahedra,
Fig.\ \ref{fig:YieldsWTime}(d), a sharp cut-off in the early yields is seen at around $T=0.13$,
corresponding to the onset of monomer starvation. Yields below this cutoff recover at longer times, as a result of the combination and rearrangement
of clusters. Similar behaviour is observed for the other shapes, 
although the sharpness of the monomer-starvation cutoff decreases for the 
smaller shapes, since monomer starvation affects smaller targets less seriously
because there are a smaller number of intermediate states in which to get 
trapped. 
Furthermore, recovery is also generally easier for the smaller targets, as
less rearrangement is likely to be required.
The virtual absence of any recovery in Fig.\ \ref{fig:YieldsWTime}(e) is 
primarily because optimal assembly for the dodecahedron occurs at a lower 
value of $\sigma_{\rm ang}$; there is some evidence of some recovery 
at larger values of $\sigma_{\rm ang}$ in Fig.\ \ref{fig:AllYields}(e).

\begin{figure}
\includegraphics[width=8.4cm]{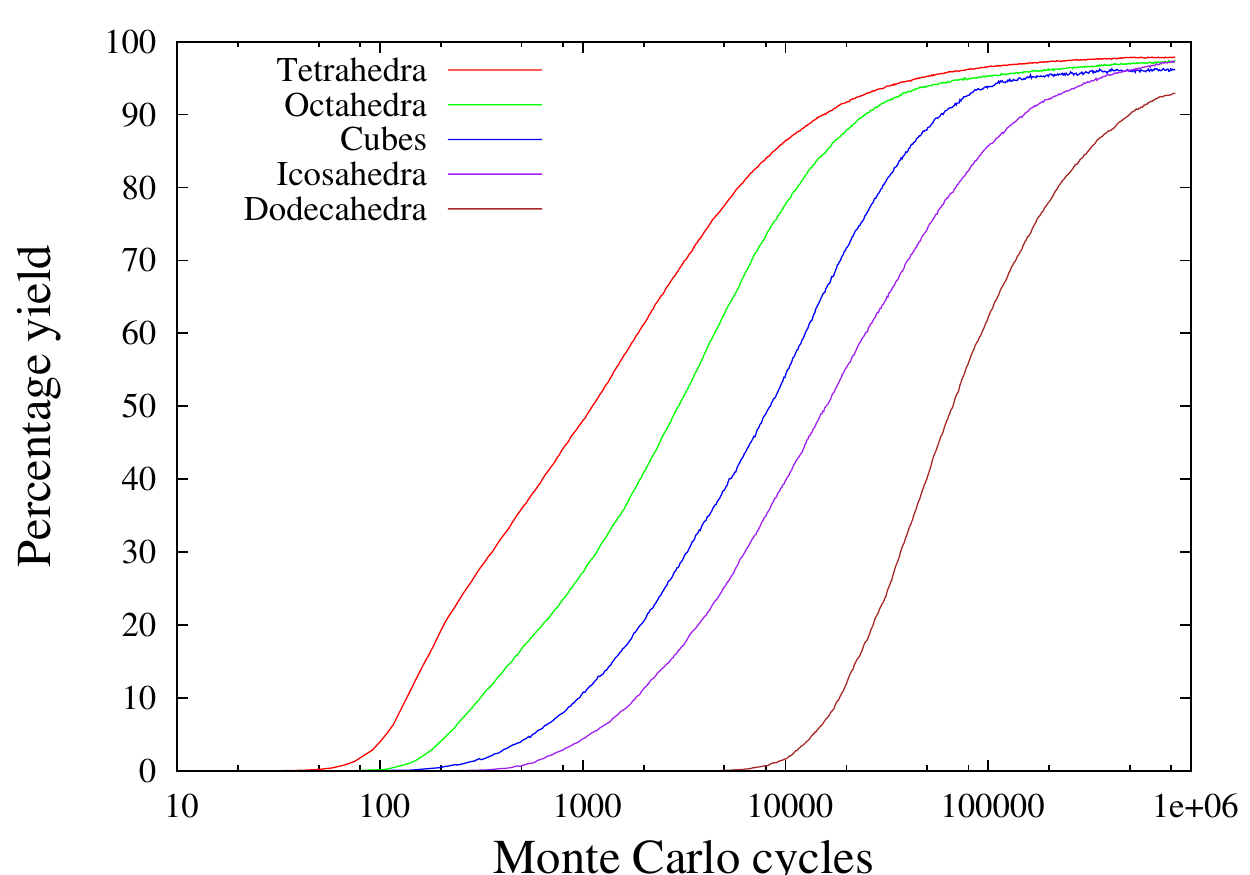}
\caption{\label{fig:YieldsVTime}(Colour Online) The yields of the various target structures as a function of time, under optimal conditions in each case.
The temperatures used were as follows. Tetrahedra: $T=0.1$, octahedra:
$T=0.12$, cubes: $T=0.09$, icosahedra: $T=0.14$, dodecahedra: $T=0.08$. $\sigma_{\rm ang} = 0.5$ for all targets except the dodecahedra, where
$\sigma_{\rm ang} = 0.35$. Each line is an average of 100 simulations, each containing 1200 particles.}
\end{figure}

Fig.\ \ref{fig:YieldsVTime} shows the yields for each target structure as a function of time. The kinetics is seen to be sigmoidal, with an initial lag
phase during which intermediates of increasing sizes build up in turn before the first complete clusters are formed. This is a ubiquitous feature in
simulations of virus assembly, and indeed predicted for all multistep reactions.\cite{FershtBook} For larger targets the lag phase is longer, since the
reaction
has to progress through a larger number of intermediates. In all cases the yield approaches $100\%$ at long times. 

The successful formation of dodecahedra is a particularly notable consequence of introducing torsional constraints. In the absence of torsional constraints
the assembly of dodecahedra is entirely prevented by competition with disordered aggregates, which are more thermodynamically stable at high
temperatures, and which rearrange only extremely slowly at lower temperatures (for more details see the accompanying paper.\cite{accompanying}) However,
since the formation
of aggregates is largely prevented by the inclusion of torsional constraints the formation of dodecahedra becomes possible, and indeed approaches $100\%$
yield at long times.
Indeed, Fig.\ \ref{fig:YieldsVTime} presents a picture of the time scale 
for assembly increasing monotonically with target size with no obviously 
anomalous behaviour due to particular geometric features of a target.

\begin{figure}
\includegraphics[width=8.4cm]{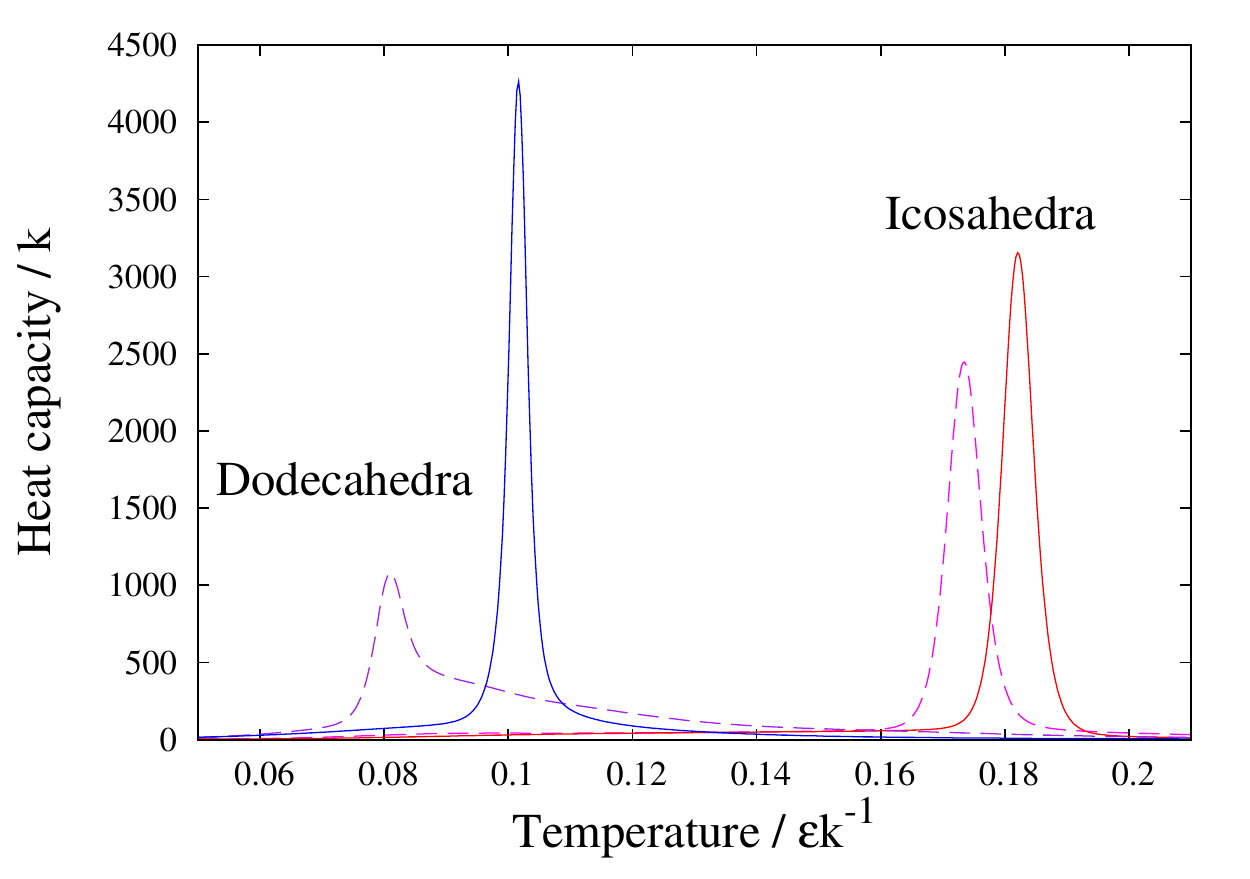}
\caption{\label{fig:IcosaAndDodecaTorCvs}(Colour Online)  The heat capacity 
$C_v$ as a function of temperature for the formation of a single icosahedron 
and dodecahedron. The dashed lines are the analogous plots for systems with
no torsional constraints.}
\end{figure}

We can see the effect of this lack of competition with aggregation in the 
thermodynamics of the assembly process. We consider first 
Fig.\ \ref{fig:IcosaAndDodecaTorCvs}, which shows the heat capacity 
$C_v$ as a function of temperature for the formation of a single icosahedron and
dodecahedron, with and without torsional constraints. 
There is a broad shoulder above the main peak of the dodecahedron
in the absence of torsional constraints,
because, as the temperature is increased, the cluster first melts, before gradually evaporating to form a monomeric vapour.  
Torsional constraints destabilise the disordered cluster state, leaving a single sharp transition between the dodecahedron and the monomer gas. 
For the icosahedron no such dramatic change
is observed, since the icosahedron directly disassembles into monomers in both 
cases. 
Rather, there is just a small increase in $T_c$ corresponding to the
slight destabilisation of the monomer gas, because it is less non-ideal.

\begin{figure}
\includegraphics[width=8.4cm]{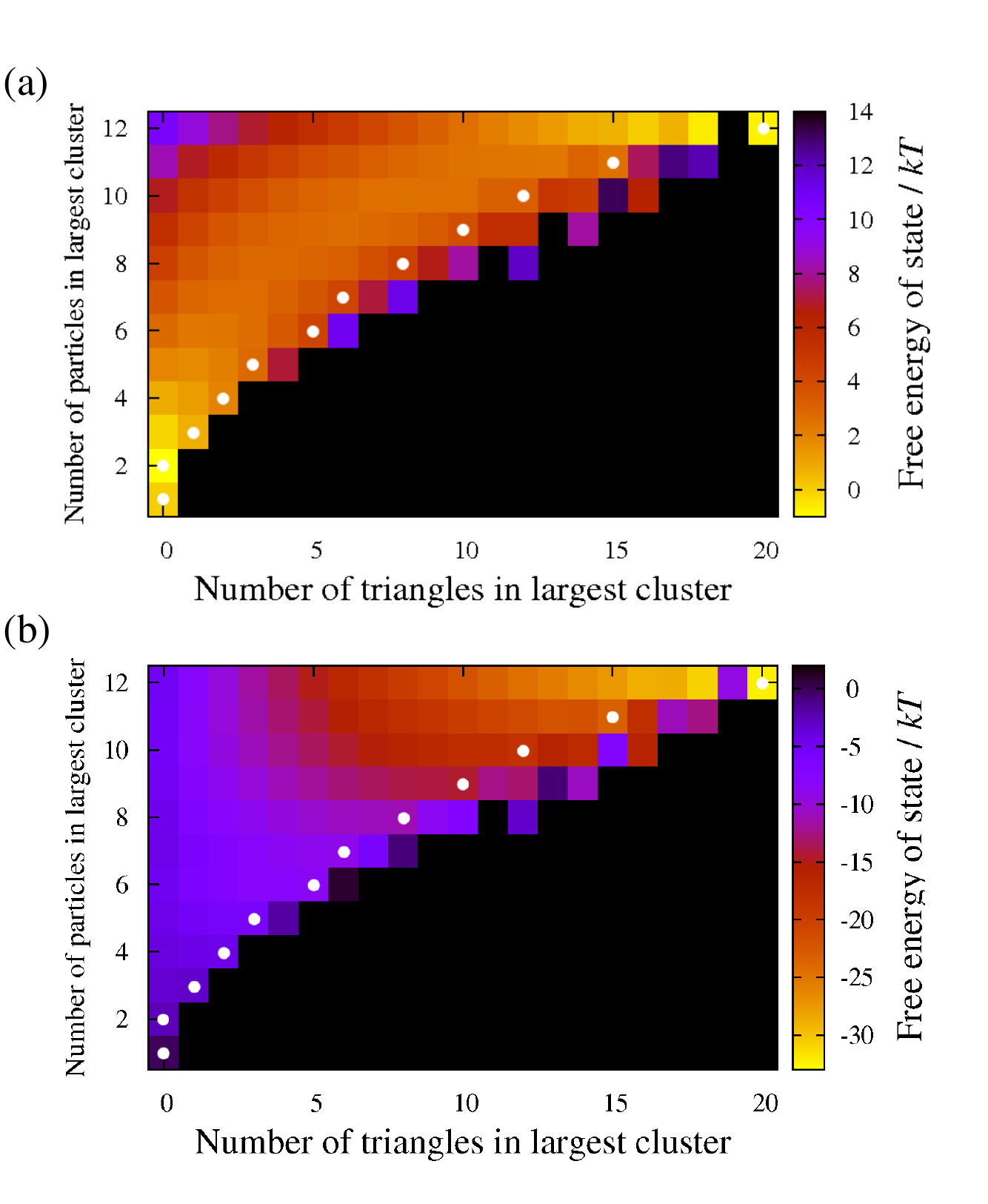}
\caption{\label{fig:IcosaLandscapes}(Colour Online) Free energy landscapes for the formation of a single icosahedron
at $\sigma_\mathrm{ang} = 0.45$ and (a) $T=0.182$, corresponding to the peak in $C_v$, (b) $T=0.16$, the optimal temperature for assembly. 
The white dots indicate sub-clusters of the icosahedron that maximize the number
of triangles for that clusters size.
}
\end{figure}

Two-dimensional free energy landscapes can provide a picture of the pathway
for self-assembly. 
Fig.\ \ref{fig:IcosaLandscapes} shows that 
there is considerable structural order (as measured by the number of triangles)
in the intermediates for icosahedron formation. This order is induced
by torsional constraints and helps the system avoid kinetic traps associated
with misformed clusters and therefore to assemble correctly.
At the optimal conditions, the most probable states for each cluster 
size are those icosahedral intermediates that maximize the number of triangles
(Fig.\ \ref{fig:IcosaLandscapes}(a)).
This result provides some support for those kinetic models that just consider
the lowest-energy pathway\cite{Endres02} or a limited set of low-energy 
pathways.\cite{Endres05,Zhang06}
However, states with fewer triangles still have significant probabilities.
Furthermore, at $T_c$ the higher entropy of states with fewer triangles makes
them more stable, and there is a broad valley across the free energy landscape,
thus implying that many pathways are relevant.

\begin{figure}
\includegraphics[width=8.0cm]{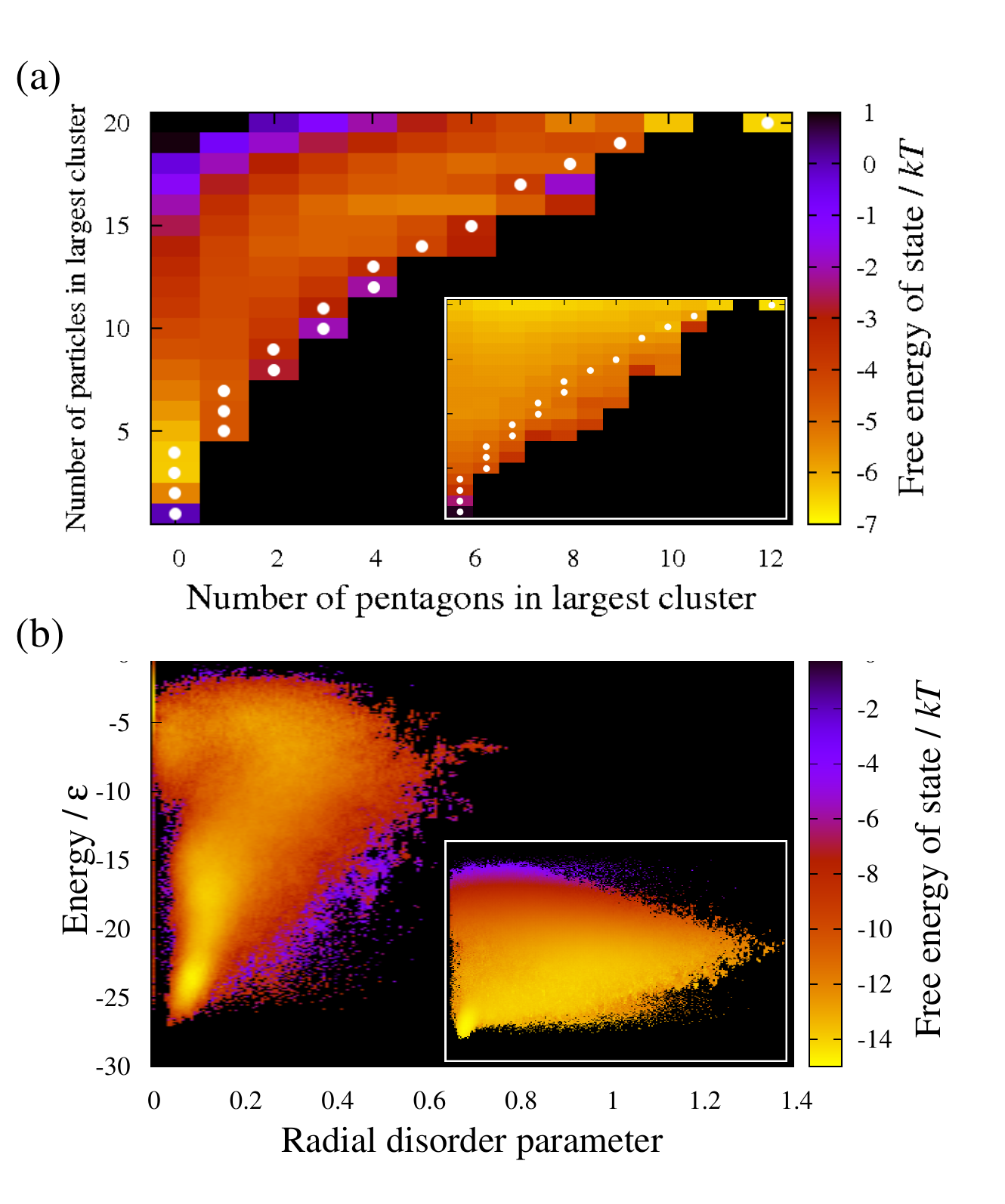}
\caption{\label{fig:DodecaLandscapes}(Colour Online) Free energy landscapes for the formation of a single dodecahedron.
In (a) the order parameters are the numbers of particles and pentagons 
in the largest cluster, while in (b) they are the potential energy and a 
measure of radial disorder, described in the text. 
The insets show the analogous plots for systems with no torsional
constraints, where the axes of the insets span the same parameter ranges 
as the main plots. 
The white dots in (a) indicate sub-clusters of the dodecahedron that maximize 
the number of pentagons for that clusters size.
}
\end{figure}

Fig.\ \ref{fig:DodecaLandscapes}(a) shows the same type of free energy 
landscape for the formation of a dodecahedron at $T_c$. 
Firstly, there are two clear free energy minima with a barrier between them, 
confirming that the transition is now between a monomeric gas and the 
assembled dodecahedron. 
Secondly, similar to the icosahedron, there is considerable structural
order along the pathway for assembly, and so as clusters grow they are `guided'
towards the dodecahedral target.
This free energy landscape contrasts sharply with the analogous landscape 
for a system without torsional constraints (see inset), 
for which disordered aggregates are also stable and block the formation of
the target structure.

A different view of the free energy landscape is provided in Fig.\ \ref{fig:DodecaLandscapes}(b) where the order parameters are the potential energy and 
a ``radial order parameter'' that is defined as the standard
deviation in the distance of particles from the centre of mass, and takes a value of zero when all particles lie on the surface of a sphere (as in a perfect
dodecahedron). 
In the absence of torsional constraints the system is able to access very
disordered states with little energy penalty, and indeed can reach low energies while remaining very far from the dodecahedral structure. 
Once torsional constraints are included the landscape takes on a classic 
funnel shape,\cite{Bryngelson95} 
such that as the energy of the system decreases it is directed into
the dodecahedral structure.

\section{Conclusions}

We have extended a patchy particle model that can self-assemble into 
monodisperse clusters\cite{Wilber07,accompanying} to 
include a torsional component in the interparticle potential, 
better mimicking the nature of protein-protein interactions. 
We found that similar principles operate in both these models. Firstly, in both
cases there is a temperature window over which successful assembly can 
occur because at too high temperatures the target structure is thermodynamically
unstable and at low temperatures the system becomes kinetically trapped.
Similarly, there is a limited range of patch specificities associated with 
successful assembly, because when the patches are too wide alternative 
configurations become possible, and when the patch-patch interactions are 
too specific growth is too slow.

However, there are also clear differences between the models. 
In particular, the torsional constraints greatly reduce the set of erroneous
structures that can be formed, and thus remove the competition with large
disordered aggregates that was a feature of the 
original model.
So now, rather than large disordered aggregates, the alternative 
configurations at larger $\sigma$
are malformed shell-like clusters, and at lower temperatures the system becomes
trapped in incomplete clusters because of monomer starvation due to the rapid
rate of cluster nucleation. Both these features make the behaviour of the
model much more similar to that for virus capsids, as would be expected from
the more protein-like nature of the interactions.

This lack of competition from aggregates enables the assembly of larger target
clusters than was possible in the original model. In particular, the 
dodecahedron which proved impossible to assemble without torsional interactions
readily assembles for the current model. The free energy landscapes for the
respective systems dramatically illustrate these differences. With torsional
constraints, the free energy landscapes now have a funnel-like topography 
guiding the system towards the target cluster, and intermediate-sized clusters
exhibit a significant amount of the structural order that is characteristic 
of the target, i.e.\ as the clusters grow, they grow with the 
correct structure.

Interestingly, the assembly behaviour for the different targets is very similar,
save for the general effects of target size, e.g.\ the time taken to achieve
assembly increases monotonically with size, and the effects of monomer 
starvation become more pronounced for larger targets. This similarity contrasts
with the original non-torsional model, where the dependence of the propensity
to aggregate on the patch geometry leads to large differences in behaviour
for the different targets. Notably, the ability of all the targets to readily
assemble for the current model suggests that proteins are less limited in the 
geometric forms into which they can assemble, as also seems apparent from 
the diverse range of biological structures and machines made up of proteins.

If synthetic particles are to begin to mimic this biological repertoire, our 
results suggest that torsionally-specific interactions between the particles
would be required. However, it is not apparent how such orientational 
specificity might be achieved for patchy nanoparticles and colloids. 
By contrast, the DNA units constructed by Mao and coworkers have both `valency'
and control over the relative orientation of the units. Thus, our results help 
to understand why tetrahedra, cubes, dodecahedra and 
truncated icosahedra can be successfully assembled from 3-armed 
units,\cite{He08,Zhang09}  and icosahedra from 5-armed units.\cite{Zhang08}  
Although far from spherical, the 
current model may not provide such a bad representation for these DNA systems, 
especially if internal degrees of freedom for the particles could be added 
to mimic the flexibility of the arms.

\begin{acknowledgments}
The authors are grateful for financial support from the EPSRC and the Royal Society.

\end{acknowledgments}


\begin{thebibliography}{66}
\expandafter\ifx\csname natexlab\endcsname\relax\def\natexlab#1{#1}\fi
\expandafter\ifx\csname bibnamefont\endcsname\relax
  \def\bibnamefont#1{#1}\fi
\expandafter\ifx\csname bibfnamefont\endcsname\relax
  \def\bibfnamefont#1{#1}\fi
\expandafter\ifx\csname citenamefont\endcsname\relax
  \def\citenamefont#1{#1}\fi
\expandafter\ifx\csname url\endcsname\relax
  \def\url#1{\texttt{#1}}\fi
\expandafter\ifx\csname urlprefix\endcsname\relax\def\urlprefix{URL }\fi
\providecommand{\bibinfo}[2]{#2}
\providecommand{\eprint}[2][]{\url{#2}}

\bibitem[{\citenamefont{Whitesides and Grzybowski}(2002)}]{Whitesides02b}
\bibinfo{author}{\bibfnamefont{G.~M.} \bibnamefont{Whitesides}}
  \bibnamefont{and}
  \bibinfo{author}{\bibfnamefont{B.}~\bibnamefont{Grzybowski}},
  \bibinfo{journal}{Science} \textbf{\bibinfo{volume}{295}},
  \bibinfo{pages}{2418} (\bibinfo{year}{2002}).

\bibitem[{\citenamefont{Goodsell}(2004)}]{Goodsell04}
\bibinfo{author}{\bibfnamefont{D.~S.} \bibnamefont{Goodsell}},
  \emph{\bibinfo{title}{Bionanotechnology}} (\bibinfo{publisher}{Wiley-Liss},
  \bibinfo{address}{Hoboken}, \bibinfo{year}{2004}).

\bibitem[{\citenamefont{Bancroft et~al.}(1967)\citenamefont{Bancroft, Hills,
  and Markham}}]{Bancroft67}
\bibinfo{author}{\bibfnamefont{J.~B.} \bibnamefont{Bancroft}},
  \bibinfo{author}{\bibfnamefont{G.~J.} \bibnamefont{Hills}}, \bibnamefont{and}
  \bibinfo{author}{\bibfnamefont{R.}~\bibnamefont{Markham}},
  \bibinfo{journal}{Virology} \textbf{\bibinfo{volume}{31}},
  \bibinfo{pages}{354} (\bibinfo{year}{1967}).

\bibitem[{\citenamefont{Prevelige et~al.}(1993)\citenamefont{Prevelige, Thomas,
  and King}}]{Prevelige93}
\bibinfo{author}{\bibfnamefont{P.~E.} \bibnamefont{Prevelige}},
  \bibinfo{author}{\bibfnamefont{D.}~\bibnamefont{Thomas}}, \bibnamefont{and}
  \bibinfo{author}{\bibfnamefont{J.}~\bibnamefont{King}},
  \bibinfo{journal}{Biophys. J.} \textbf{\bibinfo{volume}{64}},
  \bibinfo{pages}{824} (\bibinfo{year}{1993}).

\bibitem[{\citenamefont{Dokland}(1999)}]{Dokland99}
\bibinfo{author}{\bibfnamefont{T.}~\bibnamefont{Dokland}},
  \bibinfo{journal}{Cell. Mol. Life Sci.} \textbf{\bibinfo{volume}{56}},
  \bibinfo{pages}{580} (\bibinfo{year}{1999}).

\bibitem[{\citenamefont{Parent et~al.}(2005)\citenamefont{Parent, Doyle,
  Anderson, and Teschke}}]{Parent05}
\bibinfo{author}{\bibfnamefont{K.~N.} \bibnamefont{Parent}},
  \bibinfo{author}{\bibfnamefont{S.~M.} \bibnamefont{Doyle}},
  \bibinfo{author}{\bibfnamefont{E.}~\bibnamefont{Anderson}}, \bibnamefont{and}
  \bibinfo{author}{\bibfnamefont{C.~M.} \bibnamefont{Teschke}},
  \bibinfo{journal}{Virology} \textbf{\bibinfo{volume}{340}},
  \bibinfo{pages}{33} (\bibinfo{year}{2005}).

\bibitem[{\citenamefont{Parent et~al.}(2006)\citenamefont{Parent, Zlotnick, and
  Teschke}}]{Parent06}
\bibinfo{author}{\bibfnamefont{K.~N.} \bibnamefont{Parent}},
  \bibinfo{author}{\bibfnamefont{A.}~\bibnamefont{Zlotnick}}, \bibnamefont{and}
  \bibinfo{author}{\bibfnamefont{C.~M.} \bibnamefont{Teschke}},
  \bibinfo{journal}{J. Mol. Biol.} \textbf{\bibinfo{volume}{359}},
  \bibinfo{pages}{1097} (\bibinfo{year}{2006}).

\bibitem[{\citenamefont{Salunke et~al.}(1989)\citenamefont{Salunke, Caspar, and
  Garcea}}]{Salunke89}
\bibinfo{author}{\bibfnamefont{D.~M.} \bibnamefont{Salunke}},
  \bibinfo{author}{\bibfnamefont{D.~L.~D.} \bibnamefont{Caspar}},
  \bibnamefont{and} \bibinfo{author}{\bibfnamefont{R.~L.}
  \bibnamefont{Garcea}}, \bibinfo{journal}{Biophys. J.}
  \textbf{\bibinfo{volume}{56}}, \bibinfo{pages}{887} (\bibinfo{year}{1989}).

\bibitem[{\citenamefont{Rombaut et~al.}(1990)\citenamefont{Rombaut, Vrijsen,
  and Boey\'{e}}}]{Rombaut90}
\bibinfo{author}{\bibfnamefont{B.}~\bibnamefont{Rombaut}},
  \bibinfo{author}{\bibfnamefont{R.}~\bibnamefont{Vrijsen}}, \bibnamefont{and}
  \bibinfo{author}{\bibfnamefont{A.}~\bibnamefont{Boey\'{e}}},
  \bibinfo{journal}{Virology} \textbf{\bibinfo{volume}{177}},
  \bibinfo{pages}{411} (\bibinfo{year}{1990}).

\bibitem[{\citenamefont{Zlotnick et~al.}(1999)\citenamefont{Zlotnick, Johnson,
  Wingfield, Stahl, and Endres}}]{Zlotnick99}
\bibinfo{author}{\bibfnamefont{A.}~\bibnamefont{Zlotnick}},
  \bibinfo{author}{\bibfnamefont{J.~M.} \bibnamefont{Johnson}},
  \bibinfo{author}{\bibfnamefont{P.~W.} \bibnamefont{Wingfield}},
  \bibinfo{author}{\bibfnamefont{S.~J.} \bibnamefont{Stahl}}, \bibnamefont{and}
  \bibinfo{author}{\bibfnamefont{D.}~\bibnamefont{Endres}},
  \bibinfo{journal}{Biochemistry} \textbf{\bibinfo{volume}{38}},
  \bibinfo{pages}{14644} (\bibinfo{year}{1999}).

\bibitem[{\citenamefont{Zlotnick et~al.}(2000)\citenamefont{Zlotnick, Aldrich,
  Johnson, Ceres, and Young}}]{Zlotnick00}
\bibinfo{author}{\bibfnamefont{A.}~\bibnamefont{Zlotnick}},
  \bibinfo{author}{\bibfnamefont{R.}~\bibnamefont{Aldrich}},
  \bibinfo{author}{\bibfnamefont{J.~M.} \bibnamefont{Johnson}},
  \bibinfo{author}{\bibfnamefont{P.}~\bibnamefont{Ceres}}, \bibnamefont{and}
  \bibinfo{author}{\bibfnamefont{M.~J.} \bibnamefont{Young}},
  \bibinfo{journal}{Virology} \textbf{\bibinfo{volume}{277}},
  \bibinfo{pages}{450} (\bibinfo{year}{2000}).

\bibitem[{\citenamefont{Casini et~al.}(2004)\citenamefont{Casini, Graham,
  Heine, Garcea, and Wu}}]{Casini04}
\bibinfo{author}{\bibfnamefont{G.~L.} \bibnamefont{Casini}},
  \bibinfo{author}{\bibfnamefont{D.}~\bibnamefont{Graham}},
  \bibinfo{author}{\bibfnamefont{D.}~\bibnamefont{Heine}},
  \bibinfo{author}{\bibfnamefont{R.~L.} \bibnamefont{Garcea}},
  \bibnamefont{and} \bibinfo{author}{\bibfnamefont{D.~T.} \bibnamefont{Wu}},
  \bibinfo{journal}{Virology} \textbf{\bibinfo{volume}{325}},
  \bibinfo{pages}{320} (\bibinfo{year}{2004}).

\bibitem[{\citenamefont{Goodsell and Olson}(2000)}]{Goodsell00}
\bibinfo{author}{\bibfnamefont{D.~S.} \bibnamefont{Goodsell}} \bibnamefont{and}
  \bibinfo{author}{\bibfnamefont{A.~J.} \bibnamefont{Olson}},
  \bibinfo{journal}{Annu. Rev. Biophys. Biomol. Struct.}
  \textbf{\bibinfo{volume}{29}}, \bibinfo{pages}{105} (\bibinfo{year}{2000}).

\bibitem[{\citenamefont{Levy et~al.}(2006)\citenamefont{Levy, Pereira-Leal,
  Chothia, and Teichmann}}]{Levy06d}
\bibinfo{author}{\bibfnamefont{E.~D.} \bibnamefont{Levy}},
  \bibinfo{author}{\bibfnamefont{J.~B.} \bibnamefont{Pereira-Leal}},
  \bibinfo{author}{\bibfnamefont{C.}~\bibnamefont{Chothia}}, \bibnamefont{and}
  \bibinfo{author}{\bibfnamefont{S.~A.} \bibnamefont{Teichmann}},
  \bibinfo{journal}{PLoS Comput. Biol.} \textbf{\bibinfo{volume}{2}},
  \bibinfo{pages}{e155} (\bibinfo{year}{2006}).

\bibitem[{\citenamefont{Janin et~al.}(2008)\citenamefont{Janin, Bahadur, and
  Chakrabarti}}]{Janin08b}
\bibinfo{author}{\bibfnamefont{J.}~\bibnamefont{Janin}},
  \bibinfo{author}{\bibfnamefont{R.~P.} \bibnamefont{Bahadur}},
  \bibnamefont{and}
  \bibinfo{author}{\bibfnamefont{P.}~\bibnamefont{Chakrabarti}},
  \bibinfo{journal}{Q. Rev. Biophys.} \textbf{\bibinfo{volume}{41}},
  \bibinfo{pages}{133} (\bibinfo{year}{2008}).

\bibitem[{\citenamefont{Ford et~al.}(1984)\citenamefont{Ford, Harrison, Rice,
  Smith, Treffry, White, and Yariv}}]{Ford84}
\bibinfo{author}{\bibfnamefont{G.~C.} \bibnamefont{Ford}},
  \bibinfo{author}{\bibfnamefont{P.~M.} \bibnamefont{Harrison}},
  \bibinfo{author}{\bibfnamefont{D.~W.} \bibnamefont{Rice}},
  \bibinfo{author}{\bibfnamefont{J.~M.~A.} \bibnamefont{Smith}},
  \bibinfo{author}{\bibfnamefont{A.}~\bibnamefont{Treffry}},
  \bibinfo{author}{\bibfnamefont{J.~L.} \bibnamefont{White}}, \bibnamefont{and}
  \bibinfo{author}{\bibfnamefont{J.}~\bibnamefont{Yariv}},
  \bibinfo{journal}{Phil. Trans. R. Soc. B} \textbf{\bibinfo{volume}{304}},
  \bibinfo{pages}{551} (\bibinfo{year}{1984}).

\bibitem[{\citenamefont{Izard et~al.}(1999)\citenamefont{Izard, {\AE}varsson,
  Allen, Westphal, Perham, de~Kok, and Hol}}]{Izard99}
\bibinfo{author}{\bibfnamefont{T.}~\bibnamefont{Izard}},
  \bibinfo{author}{\bibfnamefont{A.}~\bibnamefont{{\AE}varsson}},
  \bibinfo{author}{\bibfnamefont{M.~D.} \bibnamefont{Allen}},
  \bibinfo{author}{\bibfnamefont{A.~H.} \bibnamefont{Westphal}},
  \bibinfo{author}{\bibfnamefont{R.~N.} \bibnamefont{Perham}},
  \bibinfo{author}{\bibfnamefont{A.}~\bibnamefont{de~Kok}}, \bibnamefont{and}
  \bibinfo{author}{\bibfnamefont{W.~G.~J.} \bibnamefont{Hol}},
  \bibinfo{journal}{Proc. Natl. Acad. Sci. USA} \textbf{\bibinfo{volume}{16}},
  \bibinfo{pages}{1240} (\bibinfo{year}{1999}).

\bibitem[{\citenamefont{Zlotnick and Stray}(2003)}]{Zlotnick03b}
\bibinfo{author}{\bibfnamefont{A.}~\bibnamefont{Zlotnick}} \bibnamefont{and}
  \bibinfo{author}{\bibfnamefont{S.~J.} \bibnamefont{Stray}},
  \bibinfo{journal}{Trends Biotechnol.} \textbf{\bibinfo{volume}{21}},
  \bibinfo{pages}{536} (\bibinfo{year}{2003}).

\bibitem[{\citenamefont{Prevelige}(1998)}]{Prevelige98}
\bibinfo{author}{\bibfnamefont{P.~E.} \bibnamefont{Prevelige}},
  \bibinfo{journal}{Trends Biotechnol.} \textbf{\bibinfo{volume}{16}},
  \bibinfo{pages}{61} (\bibinfo{year}{1998}).

\bibitem[{\citenamefont{Zlotnick et~al.}(2007)\citenamefont{Zlotnick, Lee,
  Bourne, Johnson, Domanico, and Stray}}]{Stray07}
\bibinfo{author}{\bibfnamefont{A.}~\bibnamefont{Zlotnick}},
  \bibinfo{author}{\bibfnamefont{A.}~\bibnamefont{Lee}},
  \bibinfo{author}{\bibfnamefont{C.~R.} \bibnamefont{Bourne}},
  \bibinfo{author}{\bibfnamefont{J.~M.} \bibnamefont{Johnson}},
  \bibinfo{author}{\bibfnamefont{P.~L.} \bibnamefont{Domanico}},
  \bibnamefont{and} \bibinfo{author}{\bibfnamefont{S.~J.} \bibnamefont{Stray}},
  \bibinfo{journal}{Nat. Protoc.} \textbf{\bibinfo{volume}{2}},
  \bibinfo{pages}{490} (\bibinfo{year}{2007}).

\bibitem[{\citenamefont{Cho et~al.}(2008)\citenamefont{Cho, Wang, Nie, Chen,
  and Shin}}]{Cho08}
\bibinfo{author}{\bibfnamefont{K.}~\bibnamefont{Cho}},
  \bibinfo{author}{\bibfnamefont{X.}~\bibnamefont{Wang}},
  \bibinfo{author}{\bibfnamefont{S.}~\bibnamefont{Nie}},
  \bibinfo{author}{\bibfnamefont{Z.~G.} \bibnamefont{Chen}}, \bibnamefont{and}
  \bibinfo{author}{\bibfnamefont{D.~M.} \bibnamefont{Shin}},
  \bibinfo{journal}{Clin. Cancer Res.} \textbf{\bibinfo{volume}{14}},
  \bibinfo{pages}{1310} (\bibinfo{year}{2008}).

\bibitem[{\citenamefont{Lewis et~al.}(2006)\citenamefont{Lewis, Destito,
  Zijlstra, Gonzalez, Quigley, Manchester, and Stuhlmann}}]{Lewis06}
\bibinfo{author}{\bibfnamefont{J.}~\bibnamefont{Lewis}},
  \bibinfo{author}{\bibfnamefont{G.}~\bibnamefont{Destito}},
  \bibinfo{author}{\bibfnamefont{A.}~\bibnamefont{Zijlstra}},
  \bibinfo{author}{\bibfnamefont{M.}~\bibnamefont{Gonzalez}},
  \bibinfo{author}{\bibfnamefont{J.}~\bibnamefont{Quigley}},
  \bibinfo{author}{\bibfnamefont{M.}~\bibnamefont{Manchester}},
  \bibnamefont{and}
  \bibinfo{author}{\bibfnamefont{H.}~\bibnamefont{Stuhlmann}},
  \bibinfo{journal}{Nat. Med.} \textbf{\bibinfo{volume}{12}},
  \bibinfo{pages}{354} (\bibinfo{year}{2006}).

\bibitem[{\citenamefont{Zhang}(2003)}]{Zhang03}
\bibinfo{author}{\bibfnamefont{S.~G.} \bibnamefont{Zhang}},
  \bibinfo{journal}{Nat. Biotechnol.} \textbf{\bibinfo{volume}{21}},
  \bibinfo{pages}{1171} (\bibinfo{year}{2003}).

\bibitem[{\citenamefont{Glotzer and Solomon}(2007)}]{Glotzer07b}
\bibinfo{author}{\bibfnamefont{S.~C.} \bibnamefont{Glotzer}} \bibnamefont{and}
  \bibinfo{author}{\bibfnamefont{M.}~\bibnamefont{Solomon}},
  \bibinfo{journal}{Nature Materials} \textbf{\bibinfo{volume}{6}},
  \bibinfo{pages}{557} (\bibinfo{year}{2007}).

\bibitem[{\citenamefont{Sorger et~al.}(1986)\citenamefont{Sorger, Stockley, and
  Harrison}}]{Sorger86}
\bibinfo{author}{\bibfnamefont{P.~K.} \bibnamefont{Sorger}},
  \bibinfo{author}{\bibfnamefont{P.~G.} \bibnamefont{Stockley}},
  \bibnamefont{and} \bibinfo{author}{\bibfnamefont{S.~C.}
  \bibnamefont{Harrison}}, \bibinfo{journal}{J. Mol. Biol.}
  \textbf{\bibinfo{volume}{191}}, \bibinfo{pages}{639} (\bibinfo{year}{1986}).

\bibitem[{\citenamefont{Willits et~al.}(2003)\citenamefont{Willits, Zhao,
  Olson, Baker, Zlotnick, Johnson, Douglas, and Young}}]{Willits03}
\bibinfo{author}{\bibfnamefont{D.}~\bibnamefont{Willits}},
  \bibinfo{author}{\bibfnamefont{X.}~\bibnamefont{Zhao}},
  \bibinfo{author}{\bibfnamefont{N.}~\bibnamefont{Olson}},
  \bibinfo{author}{\bibfnamefont{T.}~\bibnamefont{Baker}},
  \bibinfo{author}{\bibfnamefont{A.}~\bibnamefont{Zlotnick}},
  \bibinfo{author}{\bibfnamefont{J.}~\bibnamefont{Johnson}},
  \bibinfo{author}{\bibfnamefont{T.}~\bibnamefont{Douglas}}, \bibnamefont{and}
  \bibinfo{author}{\bibfnamefont{M.}~\bibnamefont{Young}},
  \bibinfo{journal}{Virology} \textbf{\bibinfo{volume}{306}},
  \bibinfo{pages}{280} (\bibinfo{year}{2003}).

\bibitem[{\citenamefont{Johnson et~al.}(2004)\citenamefont{Johnson, Willits,
  Young, and Zlotnick}}]{Johnson04}
\bibinfo{author}{\bibfnamefont{J.~M.} \bibnamefont{Johnson}},
  \bibinfo{author}{\bibfnamefont{D.~A.} \bibnamefont{Willits}},
  \bibinfo{author}{\bibfnamefont{M.~J.} \bibnamefont{Young}}, \bibnamefont{and}
  \bibinfo{author}{\bibfnamefont{A.}~\bibnamefont{Zlotnick}},
  \bibinfo{journal}{J. Mol. Biol.} \textbf{\bibinfo{volume}{335}},
  \bibinfo{pages}{455} (\bibinfo{year}{2004}).

\bibitem[{\citenamefont{Parent et~al.}(2007)\citenamefont{Parent, Suhanovsky,
  and Teschke}}]{Parent07}
\bibinfo{author}{\bibfnamefont{K.~N.} \bibnamefont{Parent}},
  \bibinfo{author}{\bibfnamefont{M.~M.} \bibnamefont{Suhanovsky}},
  \bibnamefont{and} \bibinfo{author}{\bibfnamefont{C.~M.}
  \bibnamefont{Teschke}}, \bibinfo{journal}{J. Mol. Bio.}
  \textbf{\bibinfo{volume}{365}}, \bibinfo{pages}{513} (\bibinfo{year}{2007}).

\bibitem[{\citenamefont{Hanslip et~al.}(2006)\citenamefont{Hanslip, Zaccai,
  Middelberg, and Falconer}}]{Hanslip06}
\bibinfo{author}{\bibfnamefont{S.~J.} \bibnamefont{Hanslip}},
  \bibinfo{author}{\bibfnamefont{N.~R.} \bibnamefont{Zaccai}},
  \bibinfo{author}{\bibfnamefont{A.~P.~J.} \bibnamefont{Middelberg}},
  \bibnamefont{and} \bibinfo{author}{\bibfnamefont{R.~J.}
  \bibnamefont{Falconer}}, \bibinfo{journal}{Biotechnol. Progr.}
  \textbf{\bibinfo{volume}{22}}, \bibinfo{pages}{554} (\bibinfo{year}{2006}).

\bibitem[{\citenamefont{Bourne et~al.}(2008)\citenamefont{Bourne, Lee,
  Venkataiah, Lee, Korba, Finn, and Zlotnick}}]{Zlotnick08b}
\bibinfo{author}{\bibfnamefont{C.}~\bibnamefont{Bourne}},
  \bibinfo{author}{\bibfnamefont{S.}~\bibnamefont{Lee}},
  \bibinfo{author}{\bibfnamefont{B.}~\bibnamefont{Venkataiah}},
  \bibinfo{author}{\bibfnamefont{A.}~\bibnamefont{Lee}},
  \bibinfo{author}{\bibfnamefont{B.}~\bibnamefont{Korba}},
  \bibinfo{author}{\bibfnamefont{M.~G.} \bibnamefont{Finn}}, \bibnamefont{and}
  \bibinfo{author}{\bibfnamefont{A.}~\bibnamefont{Zlotnick}},
  \bibinfo{journal}{J. Virol.} \textbf{\bibinfo{volume}{82}},
  \bibinfo{pages}{10262} (\bibinfo{year}{2008}).

\bibitem[{\citenamefont{Tuma et~al.}(2008)\citenamefont{Tuma, Tsuruta, French,
  and Prevelige}}]{Tuma08}
\bibinfo{author}{\bibfnamefont{R.}~\bibnamefont{Tuma}},
  \bibinfo{author}{\bibfnamefont{H.}~\bibnamefont{Tsuruta}},
  \bibinfo{author}{\bibfnamefont{K.~H.} \bibnamefont{French}},
  \bibnamefont{and} \bibinfo{author}{\bibfnamefont{P.~E.}
  \bibnamefont{Prevelige}}, \bibinfo{journal}{J. Mol. Biol.}
  \textbf{\bibinfo{volume}{381}}, \bibinfo{pages}{1395} (\bibinfo{year}{2008}).

\bibitem[{\citenamefont{Endres and Zlotnick}(2002)}]{Endres02}
\bibinfo{author}{\bibfnamefont{D.}~\bibnamefont{Endres}} \bibnamefont{and}
  \bibinfo{author}{\bibfnamefont{A.}~\bibnamefont{Zlotnick}},
  \bibinfo{journal}{Biophys. J.} \textbf{\bibinfo{volume}{83}},
  \bibinfo{pages}{1217} (\bibinfo{year}{2002}).

\bibitem[{\citenamefont{Zlotnick}(2005)}]{Zlotnick05_TheoreticalAspects}
\bibinfo{author}{\bibfnamefont{A.}~\bibnamefont{Zlotnick}},
  \bibinfo{journal}{J. Mol. Recognit.} \textbf{\bibinfo{volume}{18}},
  \bibinfo{pages}{479} (\bibinfo{year}{2005}).

\bibitem[{\citenamefont{Rapaport}(2004)}]{Rapaport04}
\bibinfo{author}{\bibfnamefont{D.~C.} \bibnamefont{Rapaport}},
  \bibinfo{journal}{Phys. Rev. E} \textbf{\bibinfo{volume}{70}},
  \bibinfo{pages}{051905} (\bibinfo{year}{2004}).

\bibitem[{\citenamefont{Rapaport}(2008)}]{Rapaport08}
\bibinfo{author}{\bibfnamefont{D.~C.} \bibnamefont{Rapaport}},
  \bibinfo{journal}{Phys. Rev. Lett.} \textbf{\bibinfo{volume}{101}},
  \bibinfo{pages}{186101} (\bibinfo{year}{2008}).

\bibitem[{\citenamefont{Endres et~al.}(2005)\citenamefont{Endres, Miyahara,
  Moisant, and Zlotnick}}]{Endres05}
\bibinfo{author}{\bibfnamefont{D.}~\bibnamefont{Endres}},
  \bibinfo{author}{\bibfnamefont{M.}~\bibnamefont{Miyahara}},
  \bibinfo{author}{\bibfnamefont{P.}~\bibnamefont{Moisant}}, \bibnamefont{and}
  \bibinfo{author}{\bibfnamefont{A.}~\bibnamefont{Zlotnick}},
  \bibinfo{journal}{Protein Sci.} \textbf{\bibinfo{volume}{14}},
  \bibinfo{pages}{1518} (\bibinfo{year}{2005}).

\bibitem[{\citenamefont{Zlotnick}(2007)}]{Zlotnick07}
\bibinfo{author}{\bibfnamefont{A.}~\bibnamefont{Zlotnick}},
  \bibinfo{journal}{J. Mol. Biol.} \textbf{\bibinfo{volume}{366}},
  \bibinfo{pages}{14} (\bibinfo{year}{2007}).

\bibitem[{\citenamefont{Zandi et~al.}(2006)\citenamefont{Zandi, van~der Schoot,
  Reguera, Kegel, and Reiss}}]{Zandi06}
\bibinfo{author}{\bibfnamefont{R.}~\bibnamefont{Zandi}},
  \bibinfo{author}{\bibfnamefont{P.}~\bibnamefont{van~der Schoot}},
  \bibinfo{author}{\bibfnamefont{D.}~\bibnamefont{Reguera}},
  \bibinfo{author}{\bibfnamefont{W.}~\bibnamefont{Kegel}}, \bibnamefont{and}
  \bibinfo{author}{\bibfnamefont{H.}~\bibnamefont{Reiss}},
  \bibinfo{journal}{Biophys. J.} \textbf{\bibinfo{volume}{90}},
  \bibinfo{pages}{1939} (\bibinfo{year}{2006}).

\bibitem[{\citenamefont{van~der Schoot and Zandi}(2007)}]{vanDerSchoot07}
\bibinfo{author}{\bibfnamefont{P.}~\bibnamefont{van~der Schoot}}
  \bibnamefont{and} \bibinfo{author}{\bibfnamefont{R.}~\bibnamefont{Zandi}},
  \bibinfo{journal}{Phys. Biol.} \textbf{\bibinfo{volume}{4}},
  \bibinfo{pages}{296} (\bibinfo{year}{2007}).

\bibitem[{\citenamefont{Twarock}(2006)}]{Twarock06}
\bibinfo{author}{\bibfnamefont{R.}~\bibnamefont{Twarock}},
  \bibinfo{journal}{Philos. Trans. R. Soc. London, Ser. A}
  \textbf{\bibinfo{volume}{364}}, \bibinfo{pages}{3357} (\bibinfo{year}{2006}).

\bibitem[{\citenamefont{Zhang and Schwartz}(2006)}]{Zhang06}
\bibinfo{author}{\bibfnamefont{T.}~\bibnamefont{Zhang}} \bibnamefont{and}
  \bibinfo{author}{\bibfnamefont{R.}~\bibnamefont{Schwartz}},
  \bibinfo{journal}{Biophys. J.} \textbf{\bibinfo{volume}{90}},
  \bibinfo{pages}{57} (\bibinfo{year}{2006}).

\bibitem[{\citenamefont{Sweeney et~al.}(2008)\citenamefont{Sweeney, Zhang, and
  Schwartz}}]{Schwartz08}
\bibinfo{author}{\bibfnamefont{B.}~\bibnamefont{Sweeney}},
  \bibinfo{author}{\bibfnamefont{T.~Q.} \bibnamefont{Zhang}}, \bibnamefont{and}
  \bibinfo{author}{\bibfnamefont{R.}~\bibnamefont{Schwartz}},
  \bibinfo{journal}{Biophys. J.} \textbf{\bibinfo{volume}{94}},
  \bibinfo{pages}{772} (\bibinfo{year}{2008}).

\bibitem[{\citenamefont{Hagan and Chandler}(2006)}]{Hagan06}
\bibinfo{author}{\bibfnamefont{M.~F.} \bibnamefont{Hagan}} \bibnamefont{and}
  \bibinfo{author}{\bibfnamefont{D.}~\bibnamefont{Chandler}},
  \bibinfo{journal}{Biophys. J.} \textbf{\bibinfo{volume}{91}},
  \bibinfo{pages}{42} (\bibinfo{year}{2006}).

\bibitem[{\citenamefont{Hagan}(2008)}]{Hagan08}
\bibinfo{author}{\bibfnamefont{M.~F.} \bibnamefont{Hagan}},
  \bibinfo{journal}{Phys. Rev. E} \textbf{\bibinfo{volume}{77}},
  \bibinfo{pages}{051904} (\bibinfo{year}{2008}).

\bibitem[{\citenamefont{Elrad and Hagan}(2008)}]{Hagan08_2}
\bibinfo{author}{\bibfnamefont{O.~M.} \bibnamefont{Elrad}} \bibnamefont{and}
  \bibinfo{author}{\bibfnamefont{M.~F.} \bibnamefont{Hagan}},
  \bibinfo{journal}{Nano Lett.} \textbf{\bibinfo{volume}{8}},
  \bibinfo{pages}{3850} (\bibinfo{year}{2008}).

\bibitem[{\citenamefont{Nguyen et~al.}(2007)\citenamefont{Nguyen, Reddy, and
  Brooks~III}}]{Brooks07}
\bibinfo{author}{\bibfnamefont{H.~D.} \bibnamefont{Nguyen}},
  \bibinfo{author}{\bibfnamefont{V.~S.} \bibnamefont{Reddy}}, \bibnamefont{and}
  \bibinfo{author}{\bibfnamefont{C.~L.} \bibnamefont{Brooks~III}},
  \bibinfo{journal}{Nano Lett.} \textbf{\bibinfo{volume}{7}},
  \bibinfo{pages}{338} (\bibinfo{year}{2007}).

\bibitem[{\citenamefont{Nguyen and Brooks~III}(2008)}]{Brooks08}
\bibinfo{author}{\bibfnamefont{H.~D.} \bibnamefont{Nguyen}} \bibnamefont{and}
  \bibinfo{author}{\bibfnamefont{C.~L.} \bibnamefont{Brooks~III}},
  \bibinfo{journal}{Nano Lett.} \textbf{\bibinfo{volume}{8}},
  \bibinfo{pages}{4574} (\bibinfo{year}{2008}).

\bibitem[{\citenamefont{Nguyen et~al.}(2009)\citenamefont{Nguyen, Reddy, and
  Brooks~III}}]{Nguyen08b}
\bibinfo{author}{\bibfnamefont{H.~D.} \bibnamefont{Nguyen}},
  \bibinfo{author}{\bibfnamefont{V.~S.} \bibnamefont{Reddy}}, \bibnamefont{and}
  \bibinfo{author}{\bibfnamefont{C.~L.} \bibnamefont{Brooks~III}},
  \bibinfo{journal}{J. Am. Chem. Soc.} \textbf{\bibinfo{volume}{131}},
  \bibinfo{pages}{2606} (\bibinfo{year}{2009}).

\bibitem[{\citenamefont{Wales}(2005)}]{Wales05}
\bibinfo{author}{\bibfnamefont{D.~J.} \bibnamefont{Wales}},
  \bibinfo{journal}{Philos. Trans. R. Soc. London, Ser. A}
  \textbf{\bibinfo{volume}{363}}, \bibinfo{pages}{357} (\bibinfo{year}{2005}).

\bibitem[{\citenamefont{Fejer et~al.}(2009)\citenamefont{Fejer, James,
  Hernandez-Rojas, and Wales}}]{Fejer09}
\bibinfo{author}{\bibfnamefont{S.~N.} \bibnamefont{Fejer}},
  \bibinfo{author}{\bibfnamefont{T.~R.} \bibnamefont{James}},
  \bibinfo{author}{\bibfnamefont{J.}~\bibnamefont{Hernandez-Rojas}},
  \bibnamefont{and} \bibinfo{author}{\bibfnamefont{D.~J.} \bibnamefont{Wales}},
  \bibinfo{journal}{Phys. Chem. Chem. Phys.} \textbf{\bibinfo{volume}{11}},
  \bibinfo{pages}{2098} (\bibinfo{year}{2009}).

\bibitem[{\citenamefont{Chen et~al.}(2007)\citenamefont{Chen, Zhang, and
  Glotzer}}]{Glotzer07}
\bibinfo{author}{\bibfnamefont{T.}~\bibnamefont{Chen}},
  \bibinfo{author}{\bibfnamefont{Z.}~\bibnamefont{Zhang}}, \bibnamefont{and}
  \bibinfo{author}{\bibfnamefont{S.~C.} \bibnamefont{Glotzer}},
  \bibinfo{journal}{P. Natl. Acad. Sci. USA} \textbf{\bibinfo{volume}{104}},
  \bibinfo{pages}{717} (\bibinfo{year}{2007}).

\bibitem[{\citenamefont{Rapaport et~al.}(1999)\citenamefont{Rapaport, Johnson,
  and Skolnick}}]{Rapaport99}
\bibinfo{author}{\bibfnamefont{D.~C.} \bibnamefont{Rapaport}},
  \bibinfo{author}{\bibfnamefont{J.~E.} \bibnamefont{Johnson}},
  \bibnamefont{and} \bibinfo{author}{\bibfnamefont{J.}~\bibnamefont{Skolnick}},
  \bibinfo{journal}{Comp. Phys. Comm.} \textbf{\bibinfo{volume}{121-122}},
  \bibinfo{pages}{232} (\bibinfo{year}{1999}).

\bibitem[{\citenamefont{Freddolino et~al.}(2006)\citenamefont{Freddolino,
  Arkhipov, Larson, McPherson, and Schulten}}]{Freddolino06}
\bibinfo{author}{\bibfnamefont{P.~L.} \bibnamefont{Freddolino}},
  \bibinfo{author}{\bibfnamefont{A.~S.} \bibnamefont{Arkhipov}},
  \bibinfo{author}{\bibfnamefont{S.~B.} \bibnamefont{Larson}},
  \bibinfo{author}{\bibfnamefont{A.}~\bibnamefont{McPherson}},
  \bibnamefont{and} \bibinfo{author}{\bibfnamefont{K.}~\bibnamefont{Schulten}},
  \bibinfo{journal}{Structure} \textbf{\bibinfo{volume}{14}},
  \bibinfo{pages}{437} (\bibinfo{year}{2006}).

\bibitem[{\citenamefont{Wilber et~al.}(2007)\citenamefont{Wilber, Doye, Louis,
  Noya, Miller, and Wong}}]{Wilber07}
\bibinfo{author}{\bibfnamefont{A.~W.} \bibnamefont{Wilber}},
  \bibinfo{author}{\bibfnamefont{J.~P.~K.} \bibnamefont{Doye}},
  \bibinfo{author}{\bibfnamefont{A.~A.} \bibnamefont{Louis}},
  \bibinfo{author}{\bibfnamefont{E.~G.} \bibnamefont{Noya}},
  \bibinfo{author}{\bibfnamefont{M.~A.} \bibnamefont{Miller}},
  \bibnamefont{and} \bibinfo{author}{\bibfnamefont{P.}~\bibnamefont{Wong}},
  \bibinfo{journal}{J. Chem. Phys.} \textbf{\bibinfo{volume}{127}},
  \bibinfo{pages}{085106} (\bibinfo{year}{2007}).

\bibitem[{\citenamefont{Wilber et~al.}(2009)\citenamefont{Wilber, Doye, and
  Louis}}]{accompanying}
\bibinfo{author}{\bibfnamefont{A.~W.} \bibnamefont{Wilber}},
  \bibinfo{author}{\bibfnamefont{J.~P.~K.} \bibnamefont{Doye}},
  \bibnamefont{and} \bibinfo{author}{\bibfnamefont{A.~A.} \bibnamefont{Louis}},
  \bibinfo{journal}{J. Chem. Phys.} \bibinfo{pages}{submitted}; arXiv.0907.4807.

\bibitem[{\citenamefont{Cho et~al.}(2007)\citenamefont{Cho, Yi, Kim, Jeon,
  Elsesser, Yu, Yang, and Pine}}]{Cho07}
\bibinfo{author}{\bibfnamefont{Y.-S.} \bibnamefont{Cho}},
  \bibinfo{author}{\bibfnamefont{G.-R.} \bibnamefont{Yi}},
  \bibinfo{author}{\bibfnamefont{S.-H.} \bibnamefont{Kim}},
  \bibinfo{author}{\bibfnamefont{S.-J.} \bibnamefont{Jeon}},
  \bibinfo{author}{\bibfnamefont{M.~T.} \bibnamefont{Elsesser}},
  \bibinfo{author}{\bibfnamefont{H.~K.} \bibnamefont{Yu}},
  \bibinfo{author}{\bibfnamefont{S.-M.} \bibnamefont{Yang}}, \bibnamefont{and}
  \bibinfo{author}{\bibfnamefont{D.~J.} \bibnamefont{Pine}},
  \bibinfo{journal}{Chem. Mater.} \textbf{\bibinfo{volume}{19}},
  \bibinfo{pages}{3183} (\bibinfo{year}{2007}).

\bibitem[{\citenamefont{He et~al.}(2008)\citenamefont{He, Ye, Zhang, Ribbe,
  Jiang, and Mao}}]{He08}
\bibinfo{author}{\bibfnamefont{Y.}~\bibnamefont{He}},
  \bibinfo{author}{\bibfnamefont{T.}~\bibnamefont{Ye}},
  \bibinfo{author}{\bibfnamefont{C.}~\bibnamefont{Zhang}},
  \bibinfo{author}{\bibfnamefont{A.~E.} \bibnamefont{Ribbe}},
  \bibinfo{author}{\bibfnamefont{W.}~\bibnamefont{Jiang}}, \bibnamefont{and}
  \bibinfo{author}{\bibfnamefont{C.}~\bibnamefont{Mao}},
  \bibinfo{journal}{Nature} \textbf{\bibinfo{volume}{452}},
  \bibinfo{pages}{198} (\bibinfo{year}{2008}).

\bibitem[{\citenamefont{Zhang et~al.}(2009)\citenamefont{Zhang, Su, He, Zhao,
  Fang, Ribbe, Jiang, and Mao}}]{Zhang09}
\bibinfo{author}{\bibfnamefont{C.}~\bibnamefont{Zhang}},
  \bibinfo{author}{\bibfnamefont{M.}~\bibnamefont{Su}},
  \bibinfo{author}{\bibfnamefont{Y.}~\bibnamefont{He}},
  \bibinfo{author}{\bibfnamefont{X.}~\bibnamefont{Zhao}},
  \bibinfo{author}{\bibfnamefont{P.-A.} \bibnamefont{Fang}},
  \bibinfo{author}{\bibfnamefont{A.~E.} \bibnamefont{Ribbe}},
  \bibinfo{author}{\bibfnamefont{W.}~\bibnamefont{Jiang}}, \bibnamefont{and}
  \bibinfo{author}{\bibfnamefont{C.}~\bibnamefont{Mao}}, \bibinfo{journal}{J.
  Am. Chem. Soc.} \textbf{\bibinfo{volume}{131}}, \bibinfo{pages}{1413}
  (\bibinfo{year}{2009}).

\bibitem[{\citenamefont{Zhang et~al.}(2008)\citenamefont{Zhang, Su, He, Zhao,
  Fang, Ribbe, Jiang, and Mao}}]{Zhang08}
\bibinfo{author}{\bibfnamefont{C.}~\bibnamefont{Zhang}},
  \bibinfo{author}{\bibfnamefont{M.}~\bibnamefont{Su}},
  \bibinfo{author}{\bibfnamefont{Y.}~\bibnamefont{He}},
  \bibinfo{author}{\bibfnamefont{X.}~\bibnamefont{Zhao}},
  \bibinfo{author}{\bibfnamefont{P.-A.} \bibnamefont{Fang}},
  \bibinfo{author}{\bibfnamefont{A.~E.} \bibnamefont{Ribbe}},
  \bibinfo{author}{\bibfnamefont{W.}~\bibnamefont{Jiang}}, \bibnamefont{and}
  \bibinfo{author}{\bibfnamefont{C.}~\bibnamefont{Mao}},
  \bibinfo{journal}{Proc. Natl. Acad. Sci. USA} \textbf{\bibinfo{volume}{105}},
  \bibinfo{pages}{10665} (\bibinfo{year}{2008}).

\bibitem[{\citenamefont{Doye et~al.}(2007)\citenamefont{Doye, Louis, Lin,
  Allen, Noya, Wilber, Kok, and Lyus}}]{Doye07}
\bibinfo{author}{\bibfnamefont{J.~P.~K.} \bibnamefont{Doye}},
  \bibinfo{author}{\bibfnamefont{A.~A.} \bibnamefont{Louis}},
  \bibinfo{author}{\bibfnamefont{I.-C.} \bibnamefont{Lin}},
  \bibinfo{author}{\bibfnamefont{L.~R.} \bibnamefont{Allen}},
  \bibinfo{author}{\bibfnamefont{E.~G.} \bibnamefont{Noya}},
  \bibinfo{author}{\bibfnamefont{A.~W.} \bibnamefont{Wilber}},
  \bibinfo{author}{\bibfnamefont{H.~C.} \bibnamefont{Kok}}, \bibnamefont{and}
  \bibinfo{author}{\bibfnamefont{R.}~\bibnamefont{Lyus}},
  \bibinfo{journal}{Phys. Chem. Chem. Phys.} \textbf{\bibinfo{volume}{9}},
  \bibinfo{pages}{2197} (\bibinfo{year}{2007}).

\bibitem[{\citenamefont{Noya et~al.}(2007)\citenamefont{Noya, Vega, Doye, and
  Louis}}]{Noya07b}
\bibinfo{author}{\bibfnamefont{E.~G.} \bibnamefont{Noya}},
  \bibinfo{author}{\bibfnamefont{C.}~\bibnamefont{Vega}},
  \bibinfo{author}{\bibfnamefont{J.~P.~K.} \bibnamefont{Doye}},
  \bibnamefont{and} \bibinfo{author}{\bibfnamefont{A.~A.} \bibnamefont{Louis}},
  \bibinfo{journal}{J. Chem. Phys.} \textbf{\bibinfo{volume}{127}},
  \bibinfo{pages}{054501} (\bibinfo{year}{2007}).

\bibitem[{\citenamefont{Villar et~al.}(2009)\citenamefont{Villar, Wilber,
  Williamson, Doye, Louis, Jochum, Lewis, and Levy}}]{Villar09}
\bibinfo{author}{\bibfnamefont{G.}~\bibnamefont{Villar}},
  \bibinfo{author}{\bibfnamefont{A.~W.} \bibnamefont{Wilber}},
  \bibinfo{author}{\bibfnamefont{P.}~\bibnamefont{Williamson},
  \bibfnamefont{A.~J.and~Thiara}}, \bibinfo{author}{\bibfnamefont{J.~P.~K.}
  \bibnamefont{Doye}}, \bibinfo{author}{\bibfnamefont{A.~A.}
  \bibnamefont{Louis}}, \bibinfo{author}{\bibfnamefont{M.~N.}
  \bibnamefont{Jochum}}, \bibinfo{author}{\bibfnamefont{A.~C.~F.}
  \bibnamefont{Lewis}}, \bibnamefont{and} \bibinfo{author}{\bibfnamefont{E.~D.}
  \bibnamefont{Levy}}, \bibinfo{journal}{Phys. Rev. Lett.}
  \textbf{\bibinfo{volume}{102}}, \bibinfo{pages}{118106}
  (\bibinfo{year}{2009}).

\bibitem[{\citenamefont{Levy et~al.}(2008)\citenamefont{Levy, Erba, Robinson,
  and Teichmann}}]{Levy08}
\bibinfo{author}{\bibfnamefont{E.~D.} \bibnamefont{Levy}},
  \bibinfo{author}{\bibfnamefont{E.~B.} \bibnamefont{Erba}},
  \bibinfo{author}{\bibfnamefont{C.~V.} \bibnamefont{Robinson}},
  \bibnamefont{and} \bibinfo{author}{\bibfnamefont{S.~A.}
  \bibnamefont{Teichmann}}, \bibinfo{journal}{Nature}
  \textbf{\bibinfo{volume}{453}}, \bibinfo{pages}{1262} (\bibinfo{year}{2008}).

\bibitem[{\citenamefont{Zlotnick}(2003)}]{Zlotnick03}
\bibinfo{author}{\bibfnamefont{A.}~\bibnamefont{Zlotnick}},
  \bibinfo{journal}{Virology} \textbf{\bibinfo{volume}{315}},
  \bibinfo{pages}{269} (\bibinfo{year}{2003}).

\bibitem[{\citenamefont{Fersht}(1999)}]{FershtBook}
\bibinfo{author}{\bibfnamefont{A.~R.} \bibnamefont{Fersht}},
  \emph{\bibinfo{title}{Structure and Mechanism in Protein Science}}
  (\bibinfo{publisher}{W.H. Freeman \& Co., New York}, \bibinfo{year}{1999}).

\bibitem[{\citenamefont{Bryngelson et~al.}(1995)\citenamefont{Bryngelson,
  Onuchic, Socci, and Wolynes}}]{Bryngelson95}
\bibinfo{author}{\bibfnamefont{J.~D.} \bibnamefont{Bryngelson}},
  \bibinfo{author}{\bibfnamefont{J.~N.} \bibnamefont{Onuchic}},
  \bibinfo{author}{\bibfnamefont{N.~D.} \bibnamefont{Socci}}, \bibnamefont{and}
  \bibinfo{author}{\bibfnamefont{P.~G.} \bibnamefont{Wolynes}},
  \bibinfo{journal}{Proteins} \textbf{\bibinfo{volume}{21}},
  \bibinfo{pages}{167} (\bibinfo{year}{1995}).

\end{thebibliography}
\end{document}